\documentclass[journal]{IEEEtran}
\usepackage{rotating,amsmath,subfigure,multirow}
\usepackage{graphicx,amssymb,lineno,bm,float}
\usepackage{algorithm,color}
\usepackage{setspace}
\usepackage{algorithmic}
\usepackage{cite}
\usepackage{array}
\usepackage{float}
\usepackage{subeqnarray}
\usepackage{cases}
\usepackage{url}
\newcommand{\tabincell}[3]{\begin{tabular}{@{}#1@{}}#2\end{tabular}}
\IEEEoverridecommandlockouts
\usepackage{booktabs}
\usepackage{longtable}

\begin{document}

\title{\huge\color{black}{A Survey on Mobile Edge Computing: The \\
Communication Perspective}}

\author{\IEEEauthorblockN{Yuyi Mao, Changsheng You, Jun Zhang, Kaibin Huang,  and Khaled B. Letaief}
\thanks{Y. Mao, J. Zhang and K. B. Letaief  are with the Dept. of Electronic and Computer Engineering, The Hong Kong University of Science and Technology, Hong Kong (Email: ymaoac@ust.hk, eejzhang@ust.hk, eekhaled@ust.hk). K. B. Letaief is also affiliated with Hamad bin Khalifa University, Doha, Qatar.

C. You and K. Huang are with the Dept. of Electrical and Electronic Engineering, The University of Hong Kong, Hong Kong (Email: csyou@eee.hku.hk,  huangkb@eee.hku.hk).
}
}

\maketitle

\begin{abstract}
Driven by the visions of Internet of Things and 5G communications, recent years have seen a paradigm shift in mobile computing, from the centralized Mobile Cloud Computing towards \emph{Mobile Edge Computing} (MEC). The main feature of MEC is to push mobile computing, network control  and storage to the network edges (e.g., base stations and access points) so as to enable  computation-intensive and latency-critical applications at the resource-limited mobile devices. MEC promises dramatic reduction in latency and mobile energy consumption,  tackling the key challenges for materializing 5G vision. The promised gains  of MEC  have  motivated extensive efforts in both academia and industry on developing the technology. A main thrust of MEC research is to seamlessly merge the two disciplines of wireless communications and mobile computing, resulting in a wide-range of new designs ranging from techniques for computation offloading to network architectures. This paper provides a comprehensive survey of the state-of-the-art MEC research with a focus on joint radio-and-computational resource management. We also discusse a set of issues, challenges and future research directions for MEC research, including  MEC system deployment, cache-enabled MEC, mobility management for MEC, green MEC, as well as privacy-aware MEC. Advancements in these directions will facilitate the transformation of MEC from theory to practice.  Finally, we introduce recent standardization efforts on MEC as well as some typical MEC application scenarios.
\end{abstract}


\IEEEpeerreviewmaketitle
\section{Introduction}
The last decade has seen Cloud Computing emerging as a new paradigm of computing. Its vision is the centralization of computing, storage and network management in the Clouds, referring to data centers, backbone IP networks and cellular core networks \cite{MArmbrust0902,zhang2010cloud}. The vast resources available in the Clouds can then be leveraged to deliver elastic computing power and storage to support resource-constrained end-user devices. Cloud Computing has been driving the rapid growth of many Internet companies. For example, the Cloud business has risen to be the most profitable sector for Amazon \cite{NYTimes16},  and Dropbox's success depended highly on the Cloud service of Amazon.

However, in recent years, a new trend in computing is happening with the function of Clouds being increasingly moving towards the network edges \cite{chiangfog}. It is estimated that tens of billions of Edge devices will be deployed in the near future, and their processor speeds are growing exponentially, following Moore's Law. Harvesting the vast amount of the idle computation power and storage space distributed at the network edges can yield sufficient capacities for performing computation-intensive and latency-critical tasks at mobile devices. This paradigm is called \emph{Mobile Edge Computing} (MEC) \cite{ETSI14}. While long propagation delays remain a key drawback for Cloud Computing, MEC, with the proximate access, is widely agreed to  be a key technology for realizing various visions for next-generation Internet, such as Tactile Internet (with millisecond-scale reaction time) \cite{Fettweis1403}, \emph{Internet of Things} (IoT) \cite{Fuqaha1506}, and Internet of Me \cite{JuniperResearch}. Presently, researchers from both academia and industry have been actively promoting MEC technology by pursuing the fusion of techniques and theories from both disciplines of \emph{mobile computing} and \emph{wireless communications}. This paper aims at providing a survey of key research progress in this young field from the communication perspective. We shall also present a research outlook containing   an ensemble  of promising research directions for MEC.

\subsection{Mobile Computing for 5G: From Clouds to Edges}
In the past decade, the popularity of mobile devices and the exponential growth of mobile Internet traffic have been driving  the tremendous advancements in wireless communications and networking. In particular,  the breakthroughs in small-cell networks, multi-antenna, and millimeter-wave  communications promise to provide users gigabit wireless access in next-generation systems \cite{andrews2014will}.  The high-rate and highly-reliable air interface allows to run computing services of mobile devices at the remote cloud data center, resulting in  the research area called \emph{Mobile Cloud Computing} (MCC). However, there is an inherent limitation of MCC, namely, the long propagation distance from the end user to the remote cloud center, which will result in excessively long latency for mobile applications. MCC is thus not adequate for a wide-range of emerging mobile applications that are latency-critical. Presently, new network architectures are being designed to better integrate the concept of Cloud Computing into mobile networks, as will be discussed in the latter part of this article.

In 5G wireless systems, ultra-dense edge devices, including  small-cell \emph{base stations} (BSs), wireless \emph{access points} (APs), laptops, tablets, and smartphones, will be deployed, each having a computation capacity comparable with that of a computer server a decade ago. As such, a large population of devices will be idle at every time instant. It will, in particular, be harvesting enormous computation and storage resources available at the network edges, which will be sufficient to enable ubiquitous mobile computing.
In a nutshell, the main target of wireless systems, from 1G to 4G, is the pursuit of increasingly higher  wireless speeds  to support the transition from voice-centric to multimedia-centric traffic. As wireless speeds approach the wireline counterparts, the mission of 5G is different and much more complex, namely to support the explosive evolution of ICT and Internet. In terms of functions, 5G systems will support \emph{communications, computing, control and content delivery} (4C). In terms of applications, a wide-range of new applications and services for 5G are emerging, such as real-time online gaming, \emph{virtual reality} (VR) and \emph{ultra-high-definition} (UHD) video streaming, which require unprecedented high access speed and low latency. The past decade also saw the take-off of different visions of next-generation Internet including IoT, Tactile Internet (with millisecond latency), Internet-of-Me, and social networks. In particular, it was  predicted by Cisco that about 50 billion IoT devices  (e.g., sensors and wearable devices) will be added to the  Internet by 2020, most of which have limited resources  for  computing, communication and storage, and have to rely on Clouds or edge devices for enhancing their capabilities\cite{evansinternet}.  It is now widely agreed that relying only on Cloud Computing is inadequate to realize the ambitious millisecond-scale latency for computing and communication in 5G. Furthermore, the data exchange between end users and remote Clouds will allow the data tsunami to saturate and bring down  the backhaul networks. This makes it essential to supplement Cloud Computing with MEC that pushes traffic, computing and network functions towards the network edges. This is also aligned with a key characteristic of next-generation networks that information is increasingly \emph{generated locally and consumed locally}, which arises from the booming of  applications in IoT, social networks and content delivery \cite{chiangfog}.

The concept of MEC was firstly proposed by the \emph{European Telecommunications Standard Institute} (ETSI) in 2014, and was defined as a new platform that ``\emph{provides IT and cloud-computing capabilities within the Radio Access Network (RAN) in close proximity to mobile subscribers}'' \cite{ETSI14}.  The original definition of MEC refers to the use of BSs for offloading computation tasks from mobile devices. Recently, the concept of \emph{Fog Computing} has been proposed by Cisco as a generalized form of MEC where the definition of edge devices gets broader, ranging from  smartphones to set-top boxes \cite{bonomi2012fog}.  This led to the emergence of a new research area called Fog Computing and Networking \cite{yi2015survey,chiangfog,klas2015fog}. However, the  areas of Fog Computing and MEC are overlapping and the terminologies are frequently used interchangeably. In this paper, we focus on MEC but many technologies discussed are also applicable to Fog Computing.

MEC is implemented based on a virtualized platform that leverages recent advancements in \emph{network functions virtualization} (NFV), \emph{information-centric networks} (ICN) and \emph{software-defined networks} (SDN). Specifically, NFV enables a single edge device to provide computing services to multiple mobile devices by creating multiple \emph{virtual machines} (VMs) \footnote{The VM is a virtual computer mapped to the physical machine's hardwares, providing virtual CPU, memory, hard drive, network interface, and other devices \cite{goldberg1974survey}. } for simultaneously performing different tasks or operating different network functions \cite{hu2015mobile}. On the other hand, ICN provides an alternative end-to-end service recognition paradigm for MEC, shifting from a host-centric to an information-centric one for implementing  context-aware computing. Last, SDN allows MEC network administrators to manage services via function abstraction, achieving scalable and dynamic computing \cite{chang2016mec}. A main focus of MEC research is to develop these general network technologies so that they can be implemented at the network edges.

\begin{figure}[!t]
\begin{center}
   \includegraphics[width=0.5\textwidth]{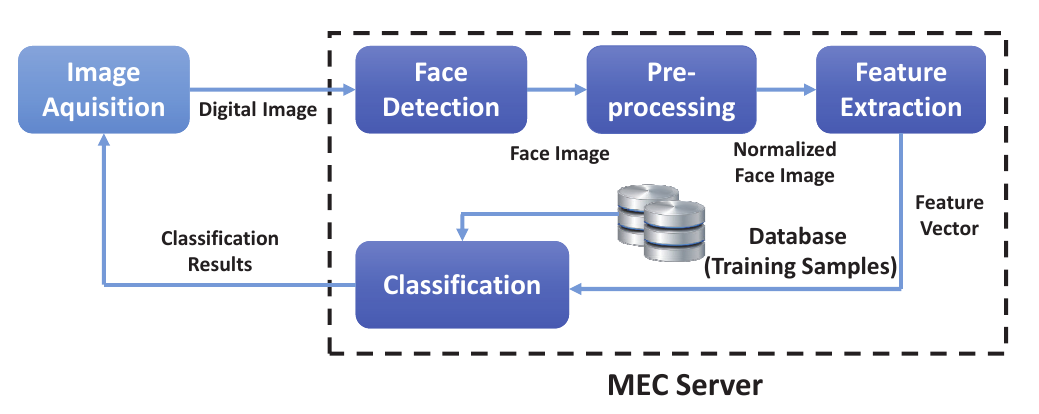}
\end{center}
\caption{Main computation components in a face recognition application \cite{Jaber1405}. }
\label{FaceRecognition}
\end{figure}

\begin{figure}[!t]
\begin{center}
   \includegraphics[width=0.4\textwidth]{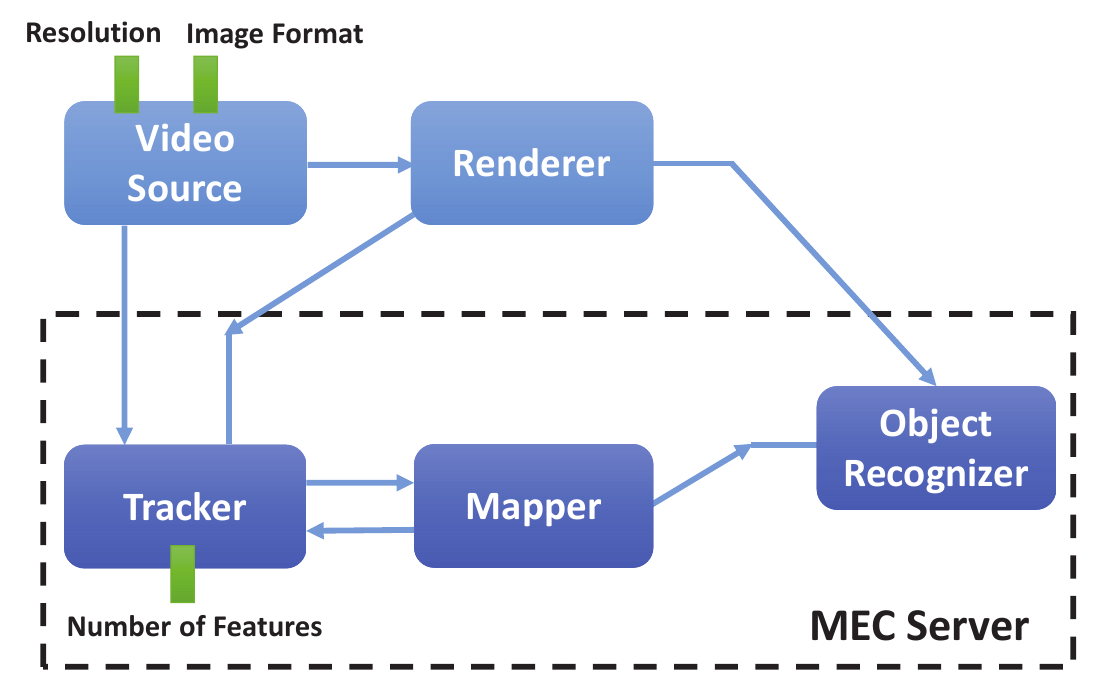}
\end{center}
\caption{Main computation components in an AR application \cite{Verbelen1310}.}
\label{ARCompo}
\end{figure}

There is an increasing number of emerging mobile applications that will benefit from MEC, by offloading their computation-intensive tasks to the MEC servers for cloud execution. In the following, we will provide two examples to illustrate the basic principles of MEC. One is the face recognition application as shown in Fig. \ref{FaceRecognition}, which typically consists of five main computation components, including image acquisition, face detection, pre-processing, feature extraction, and classification \cite{Jaber1405}. While the image acquisition component needs to be executed at the mobile device for supporting the user interface, the other components could be offloaded for cloud processing, which contain complex computation such as signal processing and \emph{machine learning} (ML) algorithms. Another popular stream of applications that can leverage the rich resources at the network edges are \emph{augmented reality} (AR) applications, which are able to combine the computer-generated data with physical reality. AR applications as shown in Fig. \ref{ARCompo} have five critical components \cite{Verbelen1310,ShuwailiAR16,ali2017energy}, namely, the video source (which obtains raw video frames from the mobile camera), a tracker (which tracks the position of the user), a mapper (which builds a model of the environment), an object recognizer (which identifies known objects in the environment), and a renderer (which prepares the processed frame for display). Among these components, the video source and renderer should be executed locally, while the most computation-intensive components, i.e., the tracker, mapper and object recognizer, can be offloaded for cloud execution. In this way, mobile users can enjoy various benefits from MEC such as latency reduction and energy savings, as will be elaborated in the next subsection.

\subsection{Mobile Edge Computing Versus Mobile Cloud Computing}

 \begin{table*}\footnotesize
  \centering
 \caption{\color{black}{Comparison of MEC and MCC Systems.}}
  \label{tab:CompMECMCC}
 \color{black}{ \begin{tabular}{ccc}
   \hline \hline
      & MEC & MCC  \\
    \hline
     \multirow{2}{*}{Server hardware}  \multirow{2}{*}  &  Small-scale data centers  & Large-scale data centers (each contains  \tabularnewline &  with moderate resources \cite{ETSI14,Juniper} &  a large number of highly-capable servers) \cite{Armbrust1004,othman2014survey} \\

     \hline
     \multirow{2}{*}{Server location}  \multirow{2}{*}  &  Co-locate with wireless gateways,  & Installed at dedicated buildings,  \tabularnewline &  WiFi routers, and LTE BSs \cite{ETSI14} & with size of several football fields \cite{BariComST13,Ghiasi1405} \\
     \hline
     \multirow{4}{*}{Deployment}  \multirow{3}{*}  &  Densely deployed by telecom operators, & Deployed by IT companies, e.g., Google   \tabularnewline  &     MEC vendors, enterprises, and & and Amazon, at a few locations \tabularnewline
       &  home users.  Require lightweight   & over the world. Require sophisticated \tabularnewline &  configuration and planning \cite{ETSI14} & configuration and planning \cite{Armbrust1004} \tabularnewline
    \hline
     \multirow{2}{*}{Distance to end users}  \multirow{2}{*}  &  Small & Large   \tabularnewline  &   (tens to hundreds of meters) \cite{hu2015mobile} & (may across the country border) \cite{Clinch1203}\tabularnewline
    \hline
    \multirow{2}{*}{Backhaul usage}  \multirow{2}{*}  &  Infrequent use & Frequent use   \tabularnewline \multirow{2}{*}  &  Alleviate congestion \cite{tran2016collaborativeNewPara} & Likely to cause congestion \cite{tran2016collaborativeNewPara} \\
    \hline
   \multirow{2}{*}{System management}  \multirow{2}{*}  & Hierarchical control & Centralized control \cite{wang2017survey} \tabularnewline \multirow{2}{*}  &  \tabincell{l}( ccentralized/distributed) \cite{wang2017survey} &   \\
    \hline
    Supportable latency & Less than tens of milliseconds \cite{hu2015mobile,JZhangChinaCom16} & Larger than 100 milliseconds \cite{CuervoMobiSys1006,satyanarayanan2009case} \tabularnewline
    \hline
     \multirow{3}{*}{Applications}  \multirow{3}{*}  &  Latency-critical and computation-intensive & Latency-tolerant and computation-intensive \tabularnewline  &     applications, e.g., AR, automatic  driving,  &  applications, e.g., online social networking,\tabularnewline & and interactive online gaming \cite{ETSI14,5GPPPauto15}.  &   and mobile commerce/health/learning \cite{KhalidTSC1412,Goel1605,Riah1509,Abbas1407}.
     \tabularnewline
     \hline
  \end{tabular}}
\end{table*}

As shown in Table~\ref{tab:CompMECMCC}, there exist significant disparities between MEC and MCC systems in terms of computing server, distance to end users and typical latency, etc.
Compared with MCC, MEC has the advantages of achieving lower latency, saving energy for mobile devices, supporting context-aware computing, and enhancing privacy and security for mobile applications. These advantages are briefly described through some examples and applications in the following.

{\bf Low Latency:} The latency for a mobile service is the aggregation of three components: \emph{propagation}, \emph{computation},   and \emph{communication} latency, depending on the propagation distance, computation capacity, and data rate, respectively. First, the information-propagation distances for MEC are typically tens-of-meters for the cases of dense small-cell networks or \emph{device-to-device} (D2D) transmissions, and typically no longer than 1km for general cases. In contrast, Cloud Computing requires transmissions from end users to nodes in core networks or data centers with distances ranging from tens of kilometers to that across continents. This results in much shorter propagation delay for MEC than that for MCC. Second,  MCC requires the information to pass through several networks including the radio-access network, backhaul network and Internet, where traffic control, routing  and other network-management operations can contribute to excessive delay. With the communication constrained at the network edges, MEC is free from  these issues.  Last, for the computation latency, a Cloud has a massive computation power that is several orders of magnitude higher than that of an edge device (e.g., a BS). However, the Cloud has to be shared by a much larger number of users than an edge device, reducing their gap in the computation latency. Furthermore, a modern BS is powerful enough for running highly sophisticated computing programs. For instance, the edge cloud at a BS has $\text{10}^\text{2}$-$\text{10}^\text{4}$ times higher computation capability than the minimum requirement (e.g., a CPU over 3.3GHz, 8GB RAM, 70GB storage space) for running the Call-of-Duty 13, a popular shooter game\footnote{\url{https://www.callofduty.com/}}. In general, experiments have shown that the total latency for MCC is in the range of 30-100ms  \cite{satyanarayanan2009case}. This is unacceptable for many latency-critical mobile applications such as real-time online gaming, virtual sports and autonomous driving, which may require tactile speed with latency approaching 1ms \cite{intelligence2014understanding}. In contrast, with short propagation distances and simple protocols, MEC has the potential of realizing tactile-level latency  for latency-critical 5G applications.

\textbf{Mobile Energy Savings:} Due to their compact forms, IoT devices have limited energy storage but are expected to cooperate and perform sophisticated tasks such as surveillance, crowd-sensing and health monitoring \cite{ASomovIEEEIoT1511}. Powering the tens of billions of IoT devices remains a key challenge for designing IoT given that frequent battery recharging/replacement is impractical if not impossible.  By effectively supporting \emph{computation offloading}, MEC stands out as a promising solution for prolonging battery lives of IoT devices. Specifically, computation-intensive tasks can be offloaded from IoT devices  to edge devices so as to reduce their energy consumption.  Significant energy savings by computation offloading have been demonstrated  in experiments, e.g., the completion of up to 44-time more computation load for  a multimedia application \emph{eyeDentify} \cite{KempEyeDentify09} or the increase of  battery life by  30-50\% for different AR applications \cite{BShiACMMulti1510}.

\textbf{Context-Awareness:} Another key feature that differentiates MEC from MCC is the ability of an MEC server for leveraging the proximity of edge devices to end users to track their  real-time information such as behaviors, locations, and environments. Inference based on such information allows the delivery of context-aware services to end users \cite{schilit1995system,perera2014context,nunna2015enabling}. For instance, the museum video guide, an AR application, can predict users' interests based on their locations in the museum to automatically deliver contents related to e.g., artworks and antiques \cite{luo2009augmented}. Another example is the CTrack system that uses the BS fingerprints to track and predict the trajectories of a large number of users for the purposes of traffic monitoring, navigation and routing, and personalized trip management \cite{thiagarajan2011accurate}.

\textbf{Privacy/Security Enhancement:} The capability of enhancing the privacy and security of mobile applications is also an attractive benefit brought by MEC compared to MCC. In MCC systems, the Cloud Computing platforms are the remote public large data centers, such as the Amazon EC2 and Microsoft Azure, which are susceptible to attacks due to their high concentration of information resources of users. In addition, the ownership and management of users' data are separated in MCC, which shall cause the issues of private data leakage and loss \cite{HSuoIWCMC13}. The use of proximate edge servers provides a promising solution to circumvent these problems. On one hand, due to the distributed deployment, small-scale nature, and the less concentration of valuable information, MEC servers are much less likely to become the target of a security attack. Second, many MEC servers could be private-owned cloudlets, which shall ease the concern of information leakage. Applications that require sensitive information exchange between end users and servers would  benefit from MEC. For instance, the enterprise deployment of MEC could help avoid uploading restricted data and material to remote data centers, as the enterprise administrator itself manages the authorization, access control, and classifies different levels of service requests without the need of an external unit \cite{ETSI_ServiceScenarios}.

\subsection{Paper Motivation and Outline}

MEC has emerged as a key enabling technology for  realizing the IoT and 5G visions \cite{coverpromise,salman2015edge,hu2015mobile}. MEC research lies at the intersection of mobile computing and wireless communications, where the existence of many  research opportunities has resulted in a highly  active area. In recent years, researchers from both academia and industry have investigated a wide-range of issues related to MEC, including system and network  modeling, optimal control, multiuser resource allocation, implementation and standardization. {\color{black}{Subsequently, several survey articles have been published to provide overviews   of the MEC area with different focuses, including system models, architectures, enabling techniques, applications, edge caching, edge computation offloading, and connections with IoT and 5G \cite{ahmed2016survey,taleb2017multi,liu2017mobile,beck2014mobile,wang2017survey,mach2017mobile,sabella2016mobile,tran2016collaborativeNewPara,ahmed2017mobile}. Their themes are summarized as follows. An overview of MEC platforms is presented in \cite{ahmed2016survey} where different  existing MEC frameworks, architectures, and their application scenarios, including FemtoClouds, REPLISM, and ME-VOLTE, are discussed. The survey of \cite{taleb2017multi} focuses on the enabling techniques in MEC such as cloud computing, VM, NFV, SDN that allow the flexible control and multi-tenancy support.  In \cite{liu2017mobile},  the authors categorize diverse MEC applications, service models, deployment scenarios, as well as network architectures. The survey in \cite{beck2014mobile} presents a taxonomy for MEC applications and identifies potential directions for research and development, such as content scaling, local connectivity, augmentation, and data aggregation and analytics.  In \cite{wang2017survey}, emerging techniques of edge \emph{computing, caching, and communications} (3C) in MEC are surveyed, showing the convergence of 3C. Besides, key enablers of MEC such as cloud technology, SDN/NFV, and smart devices are also discussed. The
survey in \cite{mach2017mobile} focuses on three critical design problems in computation offloading for MEC, namely, the offloading decision, computation resource allocation, and mobility management. In addition, the role of MEC in IoT, i.e., creating  new IoT services, is highlighted in \cite{sabella2016mobile} through MEC deployment examples with reference to IoT use cases. Several attractive use scenarios of MEC in 5G networks are also introduced in \cite{tran2016collaborativeNewPara}, ranging from mobile-edge orchestration, collaborative caching and processing, and multi-layer interference cancellation. Furthermore, potential business opportunities related to MEC are discussed in \cite{ahmed2017mobile} from the perspectives of application developers, service providers, and network equipment vendors. In view of prior work, there still lacks a systematic survey article providing comprehensive and concrete discussions on specific MEC research results with a deep integration of mobile computing and wireless communications, which motivates the current work. This paper differs from existing surveys on MEC in the following aspects. First, the current survey summarizes existing models of computing and communications in MEC to facilitate theoretical analysis and provide a quick reference for both researchers and practitioners. Next, we present a comprehensive literature review on joint radio-and-computational resource allocation for MEC, which is the central theme of the current paper. The literature review in our paper shall be a valuable addition to the existing survey literature on MEC, which can benefit readers from the research community in building up a systematic understanding of the state-of-the-art resource management techniques for MEC systems. Furthermore, we identify and discuss several research challenges and opportunities in MEC from the communication perspective, for which potential solutions are elaborated. In addition, to bridge the gap between theoretical research and real implementation of MEC, recent standardization efforts and use scenarios of MEC will then be introduced.}}
\begin{table*}\footnotesize
  \centering
  \caption{\color{black}Summary of Important Acronyms.}
  \label{tableAcron}
  {\color{black}{
  \begin{tabular}{cc|cc}
   \hline \hline
    \textbf{Acronym} & \textbf{Definition} & \textbf{Acronym} & \textbf{Definition} \\
    \hline
    AF & application function  &  MEC  &  mobile edge computing   \\
    \hline
    AR  &  augmented reality & ML &   machine learning     \\
    \hline
     AP  &  access point &   mMTC & massive machine type communication    \\
    \hline
      BS  & base station&NEF & network exposure function  \\
    \hline
       CAPEX &  capital expenditure  &  NFC &  near-filed communications   \\
    \hline
      C-RAN &  cloud radio access network & NFV &  network functions virtualization \\
    \hline
      CSI &  channel-state information & OFDMA& orthogonal frequency division multiple access   \\
    \hline
       DAG  &  directed acyclic graph  & PCF & policy control function \\
    \hline
       DCN & data-center network  &  PMR & peak-to-mean ratio  \\
    \hline
      DNS & domain name system    &  PoC & proof of concept \\
    \hline
     DP &  dynamic programming &    QoS & quality of service \\
    \hline
     DPP & determinantal point process &     RAM &  random access memory  \\
    \hline
      DVFS & dynamic frequency and voltage scaling    & RAN  &  radio access network \\
    \hline
     D2D & device-to-device  & RFID &  radio frequency identification  \\
    \hline
       EH & energy harvesting  &  RNIS & radio network information services  \\
    \hline
     eMBB & enhanced mobile broadband  & SDN &  software-defined networks  \\
    \hline
      ESI & energy side information   &  SINR & signal-to-interference-plus-noise ratio\\
    \hline
      ETSI &  European Telecommunications Standard Institute &  TOF  & traffic offloading function  \\
    \hline
      GLB & geographical load balancing  &  UE & user equipment \\
    \hline
       Het-MEC& heterogeneous MEC & UHD  &  ultra-high-definition    \\
    \hline
       HetNets & heterogeneous networks &  UPF & user plane function    \\
    \hline
       HPPP & homogeneous Poisson point process &   UPS & uninterrupted power supply     \\
    \hline
      IaaS & Infrastructure as a Service &   URLLC & ultra-reliable and low latency communication \\
    \hline
    ICN  &  information-centric networks &   VM  &  virtual machine   \\
    \hline
    ISG & industry specification group  & VR & virtual reality \\
    \hline
     ISI &  inter-symbol interference &  V2X & vehicular-to-everything \\
      \hline
     IoT  & Internet of Things   &  WPT & wireless power transfer  \\
    \hline
     KKT & Karush-Kuhn-Tucker &  3C &  computing, caching, and communications  \\
    \hline
      LP & linear programming  & 3GPP & 3rd Generation Partnership Project \\
    \hline
    LTE &  long-term evolution  &  4C & communications, computing, control and content delivery \\
    \hline
    MCC  &  mobile cloud computing  & 5GPPP & European 5G
infrastructure Public Private Partnership\\
   \hline
    MDP & Markov decision process   & 5QI & 5G QoS Indicator \\
    \hline
  \end{tabular}}}
\end{table*}

This paper is organized as follows. In Section~\ref{Section:System:Model}, we summarize the basic  MEC models, comprising models of computation tasks, communications, mobile devices   and MEC servers, based on which the models of MEC latency and energy consumption are developed. Next, a comprehensive review is presented in Section~\ref{Section:Resource:Management}, focusing on the research of joint radio-and-computational resource management for different types of MEC systems, including single-user, multiuser systems as well as multi-server MEC. Subsequently, a set of key research issues and future directions are discussed in Section~\ref{Section:Others} including 1) deployment of MEC systems, 2) cache-enabled MEC, 3) mobility management for MEC, 4) green MEC, and 5) security-and-privacy issues in MEC. Specifically, we analyze the design challenges for each research problem and provide several potential research approaches. Last, the MEC standardization efforts and applications are reviewed and discussed in Section~\ref{Section:Standard}, followed by concluding remarks in Section~\ref{Section:Conclusion}. {\color{black}We summarze the definitions of the acronyms that will be frequently use in this paper in TABLE \ref{tableAcron} for ease of reference.}

\section{MEC Computation and Communication Models} \label{Section:System:Model}
In this section, system models are introduced for the key computation/communication components of  the typical MEC system. The models provide mechanisms for abstracting various functions and operations into optimization problems and facilitating theoretical analysis as discussed in the following sections.

\begin{figure*}[!t]
\begin{center}
   \includegraphics[width=0.9\textwidth]{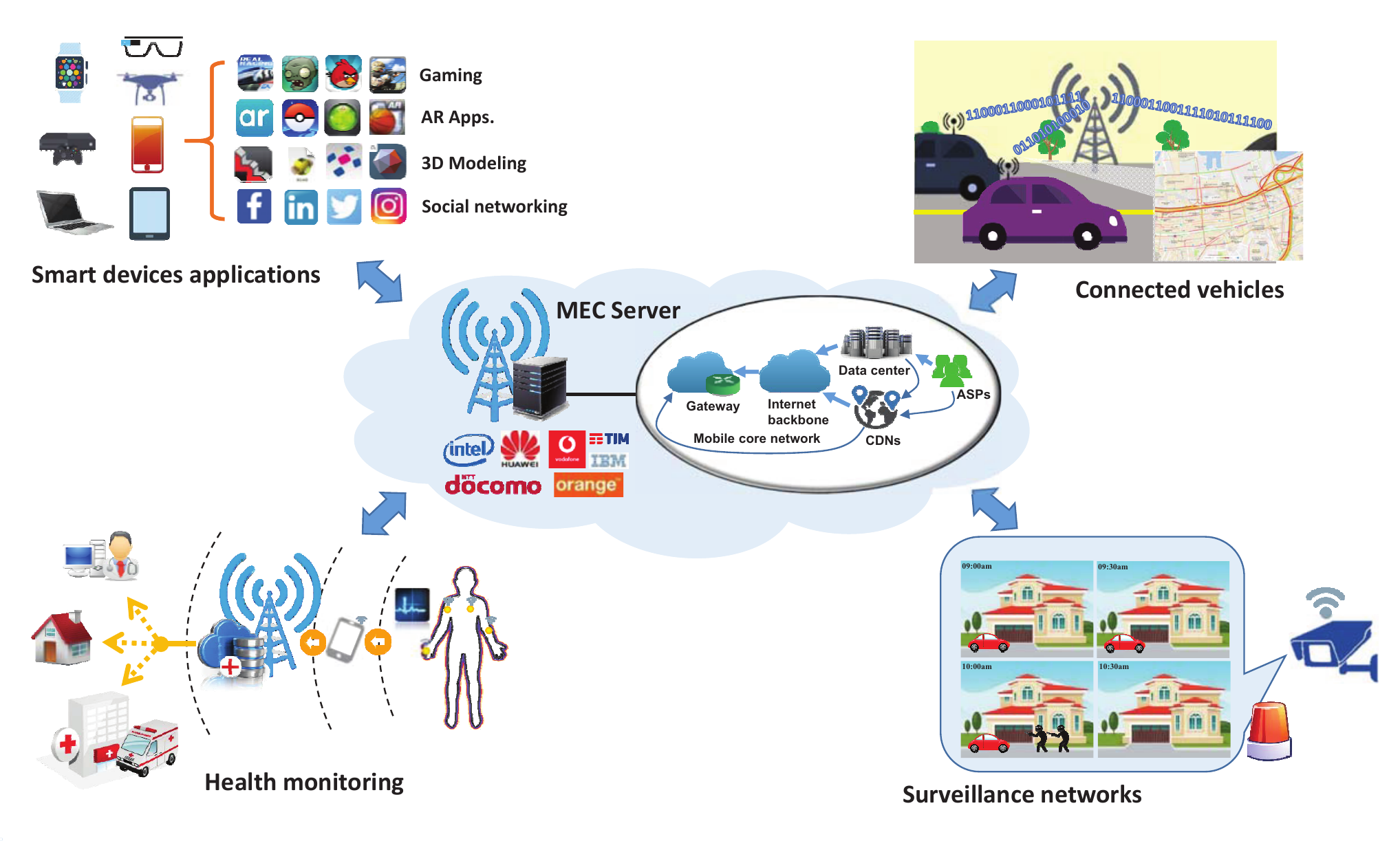}
\end{center}
\caption{Architecture of the MEC systems.}
\label{Syst_Mdl}
\end{figure*}

For the  MEC system shown in Fig. \ref{Syst_Mdl}, the key components include  mobile devices (a.k.a. end users, clients, service subscribers) and MEC servers. The MEC servers are typically small-scale data centers  deployed by the cloud and telecom operators in close proximity with end users and can be  co-located with  wireless APs. Through a gateway, the servers are connected to the  data centers via Internet. Mobile devices and servers are separated by the air interface where reliable wireless links can be established using  advanced wireless communication and networking technologies. In the following subsections, we will introduce the models for  different components of MEC systems, including models for the computation tasks, wireless communication channels and networks, as well as the computation latency and energy consumption models of mobile devices and MEC servers.

\subsection{Computation Task Models}

There are various parameters that play critical roles in modeling the computation tasks, including latency, bandwidth utilization, context awareness, generality, and scalability \cite{KhanAppModels}.  Though it is highly sophisticated to develop accurate models for tasks, there exist simple ones that are  reasonable and allow  mathematical tractability. In  this subsection, we introduce two computation-task models  popularly  used in existing literature on MCC and MEC, corresponding to  binary and partial computation offloading, respectively.

\subsubsection{\textbf{Task Model for Binary Offloading}}
A highly integrated or relatively simple task cannot be partitioned and has to be executed as a whole either  locally at  the mobile device or offloaded to the MEC server, called \emph{binary offloading}. Such a task can be  represented by a three-field notation $A\left(L,\tau_d,X\right)$. This commonly-used notation contains the information of the task input-data size $L$ (in bits), the completion deadline $\tau_{d}$ (in second), and the computation workload/intensity $X$ (in CPU cycles per bit). These parameters are related to the nature of the applications and can be estimated through  task profilers \cite{MiettinenHotCloud1006,SMelendez16}. The use of these three parameters not only captures essential properties of  mobile applications such as the computation and communication demands, but also facilitates simple evaluation of the execution latency and energy consumption performance (which will be analyzed in Section II-C).

The task $A\left(L,\tau_d,X\right)$ is required to be completed before a hard deadline $\tau_d$. This model can also be generalized to handle  the soft deadline requirement which allows a small portion of tasks to be completed after $\tau_d$ \cite{WYuan0310}. In this case, the number of CPU cycles needed to execute 1-bit of task input data is modeled as a random variable $X$. Specifically, define $x_0$ as a positive integer such that $\Pr(X >x _0) \leq \rho$ where $\rho$ is a small real number: $0 < \rho \ll 1$. It follows that $\Pr(LX > W_{\rho})\leq \rho$ where $W_{\rho} = L x_0$. Then given the $L$-bit task-input data, $W_{\rho}$ upper bounds the number of required CPU cycles almost surely.

%
\subsubsection{\textbf{Task Models for Partial Offloading}}
\begin{figure*}[!t]
\centering
\subfigure[Sequential dependency]{
\label{CG:seq}
\includegraphics[width=0.3\textwidth]{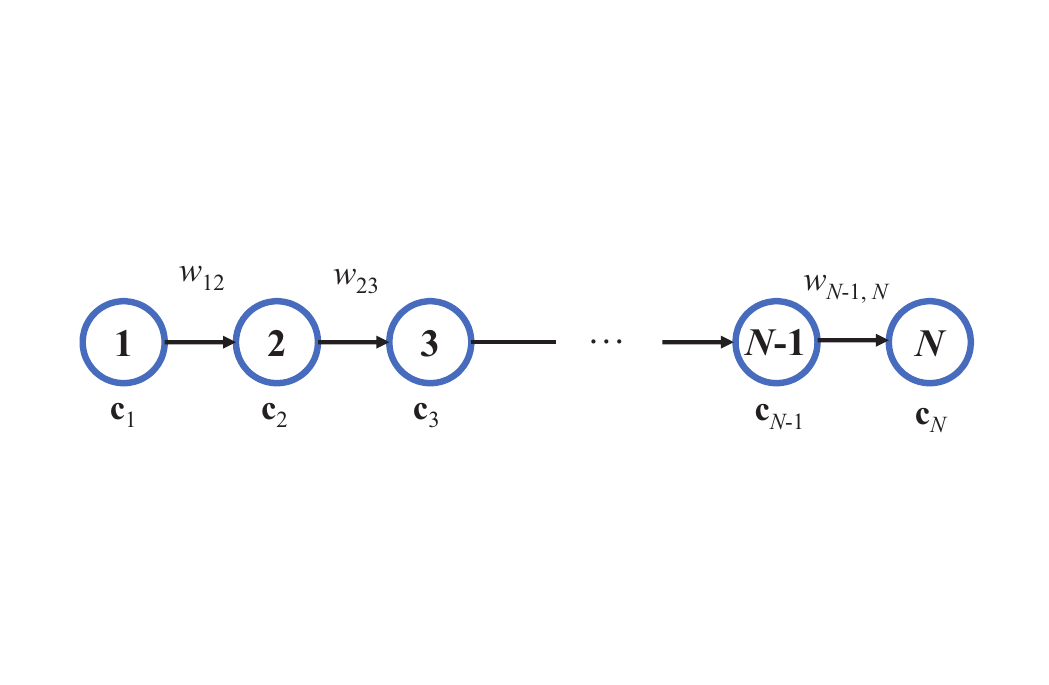}}
\subfigure[Parallel dependency]{
\label{CG:par}
\includegraphics[width=0.3\textwidth]{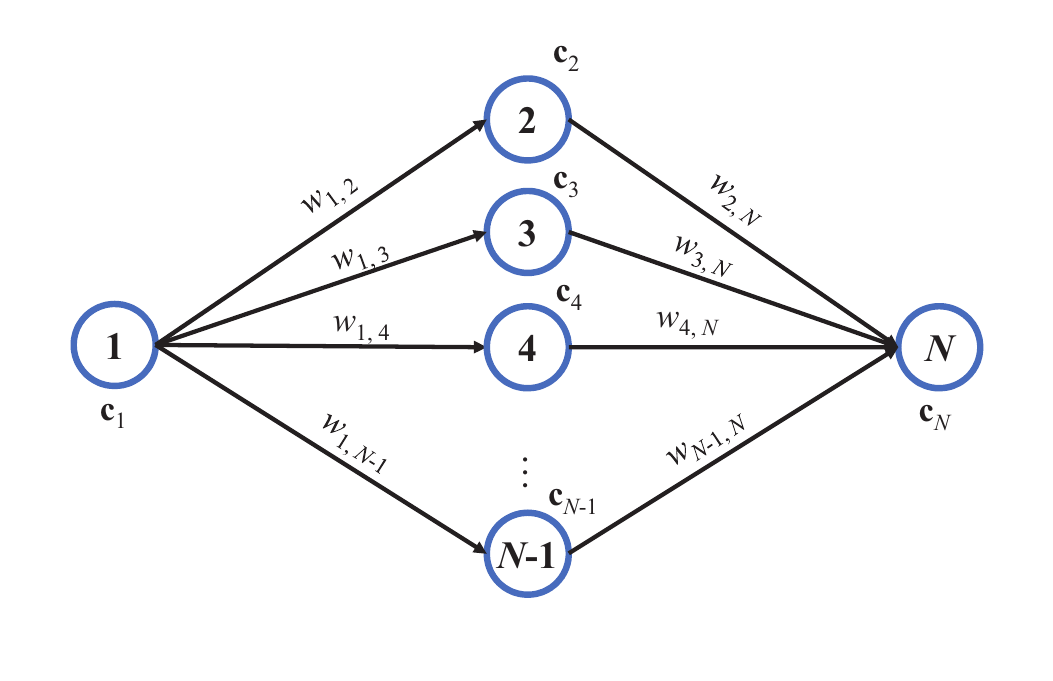}}
\subfigure[General dependency]{
\label{CG:gen}
\includegraphics[width=0.3\textwidth]{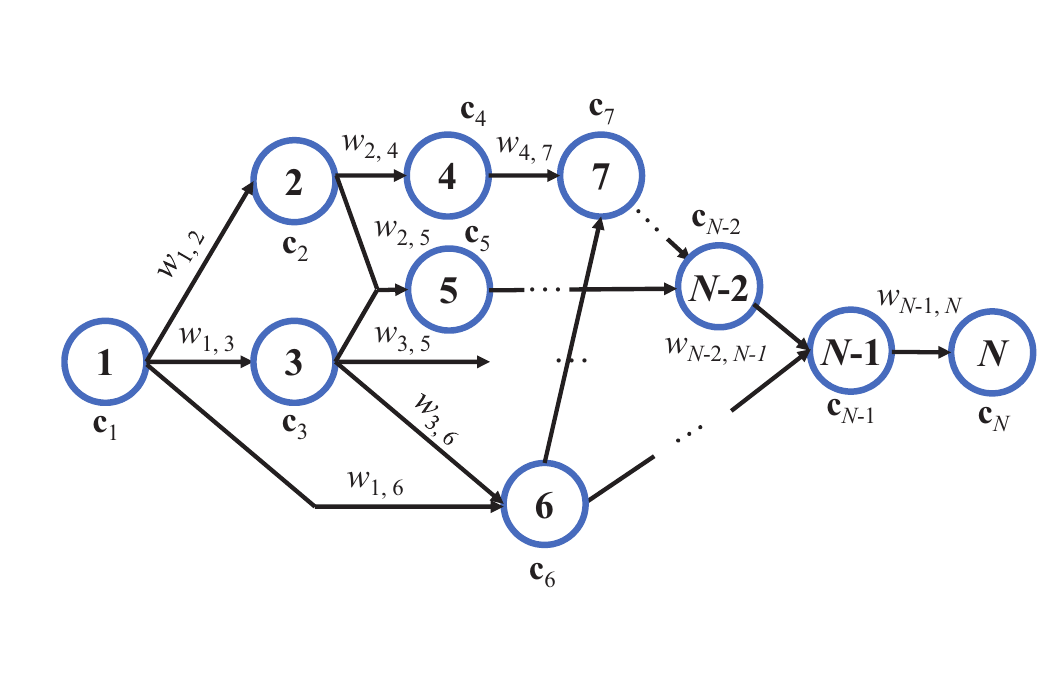}}
\caption{Typical topologies of the task-call graphs.}
\label{CGtopology}
\end{figure*}

In practice, many mobile applications are composed of multiple procedures/components (e.g., the computation components in an AR application as shown in Fig.~\ref{ARCompo}),  making it possible to implement fine-grained (partial)  computation offloading. Specifically, the program can be partitioned into two parts with one executed at the mobile device and the other  offloaded for edge execution.

The simplest task model for partial offloading is the \emph{data-partition model}, where the task-input bits are bit-wise  independent and can be arbitrarily divided into different groups and executed by different entities in MEC systems, e.g., parallel execution at the mobiles and MEC server.

Nevertheless, the dependency among different procedures/components in many applications cannot be ignored as  it  significantly affects the procedure  of execution and computation offloading due to the following reasons:
\begin{itemize}
\item First, the execution order of functions or routines  cannot be arbitrarily chosen because the outputs of some components are the inputs of others.

\item Second, due to either software or hardware constraints, some functions or routines  can be offloaded to the server for remote execution, while the  ones can only be executed locally such as the image display function.
\end{itemize}

This calls for task models that are more sophisticated than the mentioned data-partition model that can capture the inter-dependency among different computation functions and routines  in an application. One such model is called  the \emph{task-call graph}. The  graph is typically a \emph{directed acyclic graph} (DAG), which is a finite directed graph with no directed cycles. We shall denote it as $\mathcal{G}\left(\mathcal{V},\mathcal{E}\right)$, where the set of vertices $\mathcal{V}$ represents different procedures in the application and the set of edges $\mathcal{E}$ specifies their call dependencies. There are three typical dependency models of sub-tasks (i.e., task components such as functions or routines), namely \emph{sequential}, \emph{parallel}, and \emph{general}  dependency \cite{MJiaINFOCOMWS1404,MahmoodiTCC16}, as illustrated in Fig. \ref{CGtopology}. For the mobile initiated applications, the first and the last steps, e.g., collecting the I/O data and displaying the computation results on the screen, are normally required to be executed locally. Thus, node $\text{1}$ and node $N$ in Fig. \ref{CG:seq}-\ref{CG:gen} are components that must be executed locally. Besides, the required computation workloads and resources of each procedure, e.g., the number of required CPU cycles and the amount of needed memory, can also be specified in the vertices of the task-call graph, while the amount of input/output data of each procedure can be characterized by imposing weights on the edges.\\

\subsection{Communication Models}
In the literature of  MCC,  communication channels  between the mobile devices and cloud servers are typically abstracted as bit pipes with  either constant rates or random rates  with given  distributions. Such coarse  models are adopted for tractability and may be reasonable for the design of  MCC systems where the focuses are to tackle the latency in the core networks and management of large-scale cloud but not the wireless-communication latency. The scenario is different for MEC systems. Given small-scale edge clouds and targeting latency-critical applications, reducing communication latency by designing a highly efficient air interface is the main design focus. Consequently,  the mentioned bit-pipe models are insufficient as they overlook some fundamental properties of wireless propagation and are too simplified to allow the implementation of  advanced communication techniques. To be specific, wireless channels differ from the wired counterparts in the following key aspects \cite{Goldsmith2005}:
\begin{enumerate}
\item Due to  atmospheric ducting, reflection and refraction from scattering objects in the environment (e.g., buildings, walls and trees), there exists  the well-known \emph{multi-path fading} in wireless channels,  making the channels  highly time-varying and can cause severe  \emph{inter-symbol inference} (ISI). Thus, effective ISI suppression techniques, such as equalization and spread spectrum, are needed for reliable transmissions.

\item The broadcast nature of wireless transmissions results in a signal being interfered by other signals occupying the same spectrum, which reduces  their  respective receive \emph{signal-to-interference-plus-noise ratios} (SINRs) and thereby results in  the probabilities  of error in detection. To cope with the performance degradation,  interference management becomes one of the most important  design issues for  wireless communication systems and has attracted extensive research efforts \cite{GesbertMCMIMO1209,JafarTIT1401,CLiTCOM1504}.

\item Spectrum shortage has been the main foe for very high-rate radio access, motivating extensive research on exploiting new spectrum resources  \cite{TorkildsonTWC1112,XYuJSTSP1604}, designing novel transceiver architectures \cite{Alamouti9810,AGoldsmith0306,Larsson1402} and network paradigms \cite{Andrews1203,Dhillon1203} to improve the spectrum efficiency, as well as developing spectrum sharing and aggregation techniques to facilitate efficient use of fragmented and underutilized spectrum resources \cite{SHan1605,QChen1610,Kryszkiewicz1610}.
\end{enumerate}

The random variations of wireless channels in time, frequency and space make it important for designing efficient MEC systems to seamlessly integrate control of computation offloading and radio resource management.   For instance, when the wireless channel is in deep fade, the reduction on execution latency  by remote execution may not be sufficient to  compensate for the increase of  transmission latency  due to the steep drop in transmission-data  rates. For such cases,  it is desirable to defer offloading till the channel gain is favorable or switch to an alternative frequency/spatial channel  with a better quality for offloading.  Furthermore, increasing  transmission power can increase the data rate, but also lead to a larger transmission energy consumption. The above considerations necessitate the joint design of offloading and wireless transmissions, which should be adaptive to the time-varying channels based on the accurate \emph{channel-state information} (CSI).

In  MEC systems, communications are typically  between APs and mobile devices with the possibility of direct D2D communications.  The  MEC servers are small-scale data centers deployed by the Cloud Computing/telecom operators, which can be co-located with the wireless APs, e.g., the public WiFi routers and BSs, as so to reduce the \emph{capital expenditure} (CAPEX) (e.g., site rental). As shown in Fig. \ref{Syst_Mdl}, the wireless APs not only provide the wireless interface for the MEC servers, but also enable the access to the remote data center through backhaul links, which could help the MEC server to further offload some computation tasks to other MEC servers or to  large-scale cloud data centers. For the mobile devices that cannot communicate with MEC servers directly due to insufficient wireless interfaces, D2D communications with neighboring devices provide the opportunity to forward the computation tasks to MEC servers.
 Furthermore, D2D communications also enable the peer-to-peer cooperation on resource sharing and computation-load balancing within a cluster of mobile devices.

\begin{table*}\footnotesize
  \centering
  \caption{Characteristics of Typical Wireless Communication Technologies.}
  \label{tab:WCTech}
  \begin{tabular}{cccccccc}
   \hline \hline
      & NFC & RFID & Bluetooth & WiFi  & LTE & \color{black}{5G}\\
    \hline
    Max. Coverage &  10cm & 3m  & 100m & 100m  & up to 5km & \color{black}{Excellent coverage} \\
    \hline

   \multirow{3}{*}{Operation Freq.}  \multirow{3}{*}  &   & LF: 120-134kHz  &  &  & &   \tabularnewline & 13.56MHz &   HF: 13.56MHz   & 2.4GHz & 2.4GHz, 5GHz & TDD: 1.85-3.8GHz & \color{black}{6-100GHz} \tabularnewline &   & UHF: 850-960MHz   &  &  & FDD: 0.7-2.6GHz & \\ \hline

   \multirow{2}{*}{Data Rate}  \multirow{4}{*}  & & & & & & \color{black}{Indoor/dense outdoor:} \tabularnewline &  106, 212, &  Low (LF) to   &  & 135Mbps &  DL: 300Mbps & \color{black}{up to 10Gbps}\tabularnewline & 414kbps &   high (UHF)  & 22Mbps & (IEEE 802.11n) & UL: 75Mbps & \color{black}{Urban/suburban:} \tabularnewline
     & & & & & & \color{black}{$>$ hundreds of Mbps} \tabularnewline
    \hline
  \end{tabular}
\end{table*}

Presently, there exist  different types of commercialized  technologies for mobile communications, including the \emph{near-filed communications} (NFC), \emph{radio frequency identification} (RFID),  Bluetooth, WiFi, and cellular technologies such as the \emph{long-term evolution} (LTE). {\color{black}Besides, the 5G network, which will be realized by the development of LTE in combination with new radio-access technologies, is currently being standardized and will be put into commercial use as early as 2020 \cite{ericssonWhitepaper}.} These technologies can support wireless offloading from mobiles to APs or peer-to-peer mobile cooperation for varying data rates and transmission ranges.  We list the key characteristics of typical wireless communication technologies in Table~\ref{tab:WCTech}, which differ significantly in terms of the operation frequency, maximum coverage range, and data rate. For NFC, the coverage range and data rate are very low and thus the technology  is suitable for applications that require little  information exchange, e.g., \emph{e}-payment and physical access authentication. RFID is similar to NFC, but only allows one-way communications. Bluetooth is a more powerful technique to enable short-range D2D communications in MEC systems. For long-range communications between mobiles and  MEC servers, WiFi and LTE {\color{black}(or 5G in the future)} are two primary technologies enabling the access to MEC systems, which can be adaptively switched depending on their link reliability.  For the deployment of wireless technologies in MEC systems, the communication and networking protocols need to be redesigned to integrate both the computing and  communication infrastructures, and effectively improve the computation efficiency that is more sophisticated than the data transmission.


\subsection{Computation Models of Mobile Devices}\label{Sec:ModelMD}

In this subsection, we introduce the computation models of mobile devices and discuss methodologies of evaluating the computation performance.

The CPU of a mobile device  is the primary  engine  for local computation.  The CPU performance is controlled  by the CPU-cycle frequency $f_{m}$ (also known as the CPU clock speed).  The state-of-the-art mobile CPU architecture adopts the advanced \emph{dynamic frequency and voltage scaling} (DVFS) technique, which allows stepping-up or -down of the CPU-cycle frequency (or voltage), resulting in growing and reducing energy  consumption, respectively. In practice, the value of $f_{m}$ is bounded by a maximum value, $f_{\rm{CPU}}^{\max}$, which reflects the limitation of the mobile's  computation capability. Based on the computation task model introduced in Section II-A, the execution latency for task $A\left(L,\tau,X\right)$ can be calculated accordingly to
\begin{equation}
t_{m} = \frac{LX}{f_{m}},
\label{SecII:CPUlatency}
\end{equation}
which indicates that a high CPU clock speed is desirable in order to reduce the execution latency, at the cost of higher CPU energy consumption.

As the mobile devices are energy-constrained, the  energy consumption for local computation is another critical measurement for the  mobile computing efficiency. According to the circuit theory \cite{BurdVLSI96,WYuanACMTOCS0608,Vogeleer2013CPUenergy,WZhangTWC1312}, the CPU power consumption can be divided into several factors including the \emph{dynamic}, \emph{short-circuit}, and \emph{leakage} power consumption\footnote{The dynamic power consumption comes from the toggling activities of the logic gates inside a CPU, which shall charge/discharge the capacitors inside the logic gates. When a logic gate toggles, some of its transistors may change states, and thus, there might be a short period of time when some transistors are conducting simultaneously. In this case, the direct path between the source and ground will result in some short-circuit power loss. The leakage power dissipation is due to the flowing current between doped parts of the transistors \cite{Vogeleer2013CPUenergy}, available on \url{https://en.wikipedia.org/wiki/CPU_power_dissipation}.}, where the dynamic power consumption  dominates the others. In particular, it is shown in \cite{Vogeleer2013CPUenergy} that the dynamic power consumption is proportional to the product of $V_{\rm{cir}}^{2}f_{m}$ where $V_{\rm{cir}}$ is the circuit supplied voltage. It is further noticed in \cite{BurdVLSI96,WZhangTWC1312} that, the clock frequency of the CPU chip is approximately linear proportional to the voltage supply when operating at the low voltage limits. Thus, the energy consumption of a CPU cycle is given by $\kappa f^{2}_{m}$, where $\kappa$ is a constant related to the hardware architecture. For the computation task $A\left(L,\tau,X\right)$ with CPU clock speed $f_{m}$, the energy consumption can be derived:
\begin{equation}
E_{m} = \kappa LX f^{2}_{m}.
\label{SecII:CPUenergy}
\end{equation}
One can observe  from (\ref{SecII:CPUlatency}) and (\ref{SecII:CPUenergy}) that the mobile device may not be able to complete a computation-intensive task within the required  deadline, or else the energy consumption incurred by mobile execution is so high that the onboard battery will be depleted quickly. In such cases, offloading the task execution process to an MEC server is desirable.

Besides CPUs, other hardware components in the mobile devices, e.g., the \emph{random access memory} (RAM) and flash memory, also contribute to  the computation  latency and energy consumption \cite{Carroll:2010:APC:1855840.1855861}, while detailed discussions are beyond the scope of this survey.\\
\subsection{Computation Models of MEC Servers}

In this subsection, we introduce the computation models of the MEC servers. Similar as the mobile devices, the computation latency  and energy consumption are of particular interests.

The server-computation latency  is \emph{negligible} compared with communication or local-computation latency in  MEC systems where  the computation loads for servers are much lower than  their computation capacities  \cite{WZhangTWC1312, you2016energyJSAC}. This model can be also relevant for  multiuser MEC systems with resource-constrained servers  if the servers' computation loads are regulated by multiuser resource management under latency and computation-capacity constraints \cite{you2016energy}.

On the other hand, as edge servers have relatively limited computation resources, it is necessary to consider the \emph{non-negligible} server execution time in the general design of MEC systems, yielding the computation model for the severs discussed in the remainder of this subsection. Two possible models  are considered  in the literature, corresponding to  the \emph{deterministic} and \emph{stochastic} server-computation latency. The deterministic model is proposed to consider the exact server-computation latency  for latency-sensitive applications, which is implemented using  techniques such as VMs and DVFS.
Specifically, assume the MEC server allocates different VMs for different mobile devices, allowing independent computation \cite{barham2003xen}. Let $f_{s,k}$ denote the allocated servers' CPU-cycle number for mobile device $k$. Similar to Section~\ref{Sec:ModelMD}, it follows that the server execution time denoted by $t_{s,k}$ can be calculated as $t_{s,k} = \dfrac{w_k}{f_{{s,k}}}$, where $w_k$ is the number of required CPU cycles for processing the offloaded computation workload. This model has been widely used for designing computation-resource allocation policies  \cite{barbarossa2013joint,chen2015efficient,lyumulti:2016:ProxiCloud}. A similar model was proposed in \cite{you2016energy}, where the MEC server is assumed to perform load balancing for the total offloaded computation workloads. In other words, the CPU cycles at the MEC server are proportionally allocated to each mobile device such that they experience the same execution latency. Furthermore, in addition to the CPU processing time, the server scheduling queuing delay should be accounted for MEC servers with relatively small computation capacities, where parallel computing via virtualization techniques is not feasible and thus it needs to process the computation workloads sequentially. Without loss of generality, denote $k$ as the processing order for a mobile device and name it as mobile $k$. Thus, the total server-computation latency including the queuing delay for device $k$ denoted by $T_{s,k}$ can be given as
\begin{equation}
T_{s,k}=\sum_{i\leq k} t_{s,i}.
\end{equation}
For latency-tolerant applications, the average server-computation time can be derived based on stochastic models. For example, in \cite{vakilinia2015modeling}, the task arrivals and service time are modeled by the Poisson and exponential processes, respectively. Thus, the average server-computation time can be derived using techniques from queuing theory. Last, for all above models, as investigated in \cite{MArmbrust0902}, multiple VMs sharing the same physical machine will introduce the I/O interference among different VMs. It results in the longer computation latency for each VM denoted by $T_{s,k}^{'}$, which can be modeled by $T_{s,k}^{'}=T_{s,k} (1+\epsilon)^n$ where $\epsilon$ is the performance degradation factor as the percentage increasing of the latency \cite{bruneo2014stochastic}.


The energy consumption of an MEC server is jointly determined by the usage of the CPU, storage, memory, and network interfaces. Since the CPU contribution is dominant among these factors,  it is the main focus in the literature.  Two tractable models are widely used for  the energy consumption of  MEC servers. One model is based on the DVFS technique described as follows. Consider an MEC server that handles $K$ computation tasks and the $k$-th task is allocated with $w_k$ CPU cycles with CPU-cycle frequency $f_{s,k}$. Hence, the total energy consumed by the CPU at the MEC server, denoted by $E_s$, can be expressed as
\begin{equation}
E_s=\sum_{k=1}^K \kappa w_k f^{2}_{{s,k}},
\end{equation}
which is similar to that for the mobile devices. The other model is based on an observation in recent works \cite{fan2007power,lin2011energy,beloglazov2012energy} that the server-energy consumption is \emph{linear} to the CPU utilization ratio which depends on the computation load. Moreover, even for an idle server, it still, on average, consumes up to 70\% of the energy consumption for the case with  the full CPU speed. Thus, the energy consumption at the MEC server can be calculated according to
\begin{equation}
E_s=\alpha E_{\max}+(1-\alpha)E_{\max} u,
\label{ServerEnergyUtiRatio}
\end{equation}
where $E_{\max}$ is the energy consumption for a fully-utilized server, $\alpha$ is the fraction of the idle energy consumption (e.g., 70\%) and $u$ denotes the CPU utilization ratio. This model suggests that energy-efficient MEC should allow  servers to be switched into the  sleep mode in the case of  light load and consolidation of computation loads into fewer active servers.
{\color{black}{\subsection{Summary and Insights}
\begin{figure*}[!t]
\begin{center}
   \includegraphics[width=0.85\textwidth]{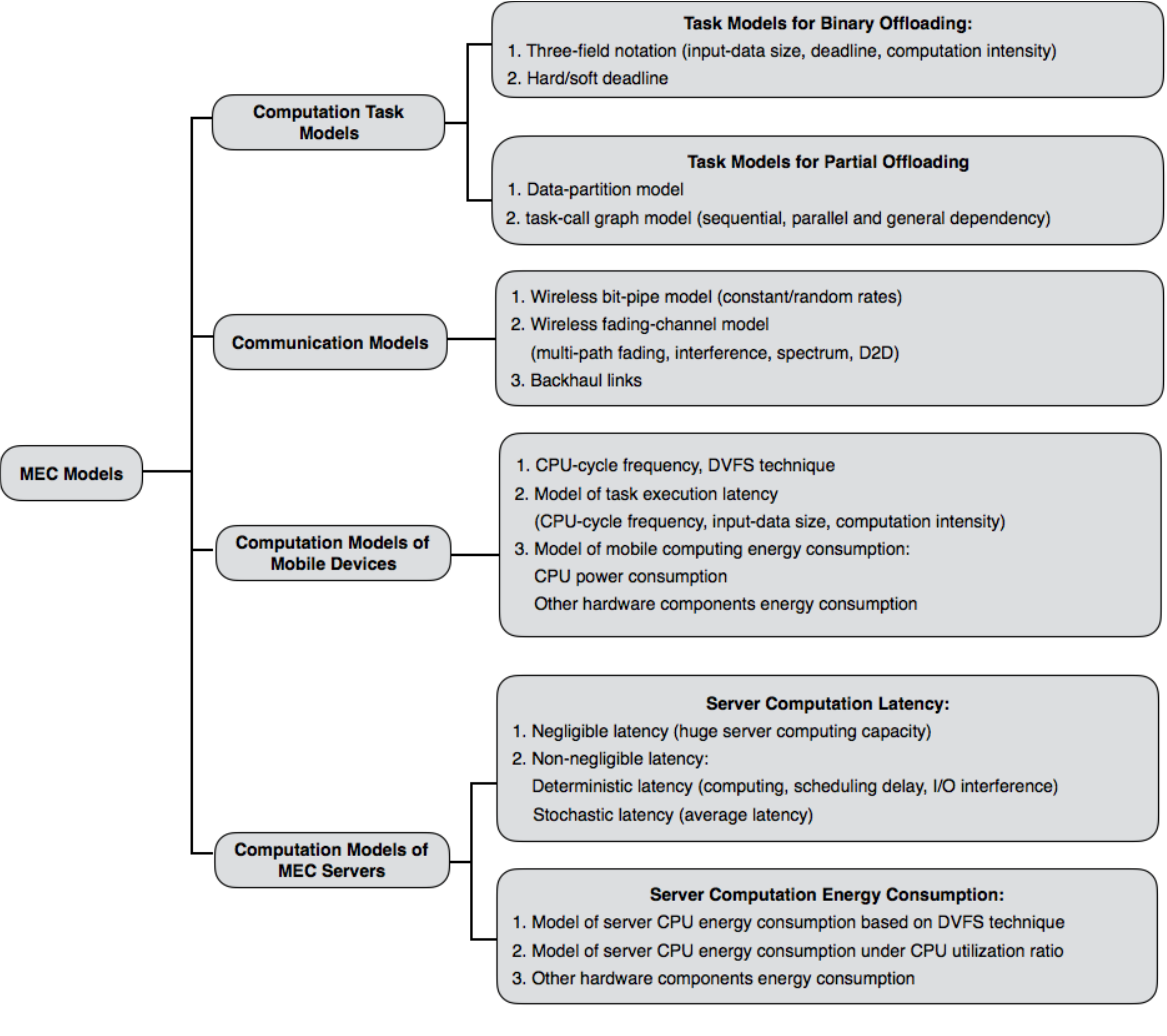}
\end{center}
\caption{\color{black}{Summary of MEC models.}}
\label{ModelSummary}
\end{figure*}
The MEC computation and communication models are summarized in Fig.~\ref{ModelSummary}, laying the foundation for the analysis of MEC resource management in the next section. These models shed several useful insights on the offloading design, listed as follows.
\begin{itemize}
\item The effective design of MEC should leverage and integrate advanced techniques from both areas of wireless communications and mobile computing.
\item It is vital to choose suitable computation task models for different MEC applications. For example, the soft-deadline task model can be applied for social networking applications but is not suitable for AR applications due to the stringent computation latency requirements. Moreover, for a specific application, the task model also depends on the offloading scenario, e.g., the data-partition model can be used when the input-data is offloaded, and the task-call graph should be considered when each task component can be offloaded as a whole.
\item The wireless channel condition significantly affects the amount of energy consumption for   computation offloading. MEC has the potential to reduce the transmission energy consumption due to short distances between users and MEC servers. Advanced wireless communication techniques, such as interference cancelation and adaptive power control, can further reduce the offloading energy consumption.
\item Dynamic CPU-cycle frequency control is the key technique for controlling the computation latency and energy consumption for both mobile devices and MEC servers. Specifically, increasing the CPU-cycle frequency can reduce the computing time but contributes to higher energy consumption. The effective CPU-cycle frequency control should approach the optimal tradeoff between computation latency and energy consumption.
\item Apart from the task-execution latency, the computation scheduling delay is non-negligible if the MEC server has a relatively small computation capacity or heavy computation loads are offloaded to the server. Load-balancing and intelligent scheduling policies can be designed to reduce the total computation latency.
\end{itemize} }}
\section{Resource Management in MEC Systems}\label{Section:Resource:Management}
\begin{figure*}[!t]
\begin{center}
   \includegraphics[width=0.8\textwidth]{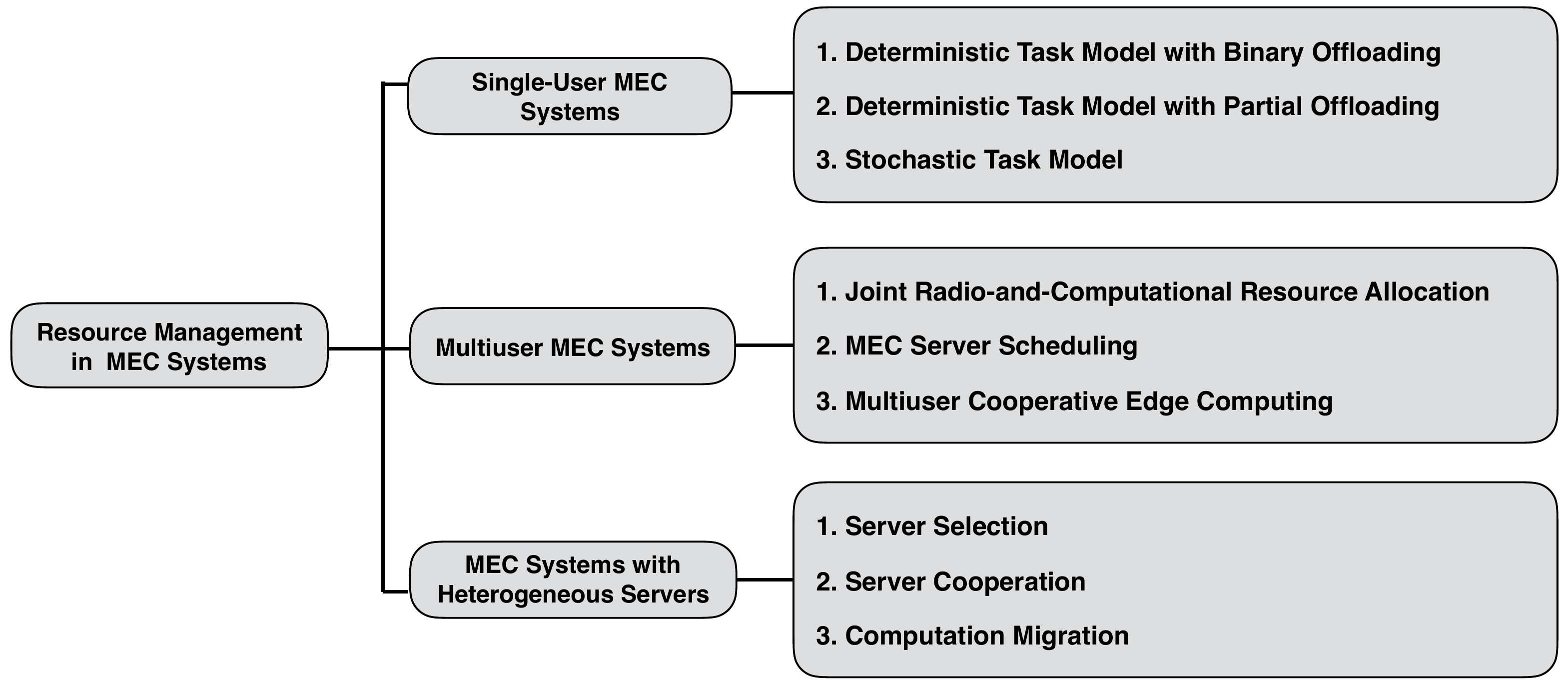}
\end{center}
\caption{Classification of resource management techniques for MEC.}
\label{SurveyTree}
\end{figure*}

The joint radio-and-computational resource management plays a pivotal role in realizing energy-efficient and low-latency MEC. The implementation of relevant techniques is facilitated by the network architecture where MEC servers and wireless APs (e.g., BSs and WiFi routers) are co-located.  In this section, we provide a comprehensive overview of the  literature on resource management  for MEC systems summarized in Fig.~\ref{SurveyTree}. Our discussion starts from the simple single-user systems comprising  a single mobile device and a single MEC server, allowing the exposition of  the key design considerations and basic design methodologies. Subsequently,  more complex multiuser  MEC systems are considered where multiple offloading users compete for the use of both the  radio and server-computation  resources and have been coordinated. Last, we  extend the discussion to MEC systems with heterogeneous  servers which not only provide the freedom of server selection but also allow  the cooperation among  servers. Such network-level  operations can  significantly enhance the performance of MEC systems.

\subsection{Single-User MEC Systems}
This subsection focuses on the simple single-user MEC systems and reviews a set of recent research efforts for this case. The discussion is divided according to three  popularly-used task models, namely, deterministic task model with binary offloading, deterministic task model with partial offloading,  and stochastic task model.
\subsubsection{\textbf{Deterministic Task Model with Binary Offloading}} Consider the mentioned  single-user MEC system where the binary offloading decision is on whether a particular task should be offloaded for edge execution or local computation.   The investigations for the optimal  offloading policies can be dated back to  those for conventional Cloud Computing systems, where the communication links were typically assumed to have a fixed rate  $B$. In  \cite{KumarCompt1004} and \cite{KumarMPA1004}, general guidelines are developed for determining the offloading decision for the purposes of minimizing the mobile-energy consumption and computation  latency. Denote $w$ as the amount of computation (in CPU cycles) for a task, $f_{m}$ as the CPU speed of the mobile device, $d$ as the input data size, and $f_{s}$ as the CPU speed at the cloud server. Offloading the computation to the cloud server can improve the latency performance only when
\begin{equation}
\frac{w}{f_{m}}> \frac{d}{B} + \frac{w}{f_{s}},
\label{SecIII:OffloadingLatency}
\end{equation}
which holds for applications that require heavy computation and have small amount of data input, or when the cloud server is fast, and the transmission rate  is sufficiently high. Moreover, let  $p_{m}$ represent the CPU power consumption at the mobile device, and $p_{t}$ as the transmission  power, $p_{i}$ as the power consumption at the device when the task is running at the server. Offloading the task could help save mobile energy when
\begin{equation}
p_{m}\times \frac{w}{f_{m}} > p_{t}\times \frac{d}{B} + p_{i} \times \frac{w}{f_{s}}
\label{SecIII:OffloadingEnergy}
\end{equation}
holds, i.e., applications with heavy computation and light communication should be offloaded.

Nevertheless, the data  rates for wireless communications are not constant and change with the time-varying channel gains as well as depend on the transmission power. This calls  for the design of control policies for  power adaptation and data scheduling to streamline the offloading process. In addition, as the CPU power consumption increases super-linearly with the CPU-cycle frequency, the computation energy consumption for mobile execution can be minimized using DVFS techniques. These issues led to the active field of adaptive MEC as summarized below.

In \cite{BarbarossaMSP1411}, the problem of transmission-energy minimization  under a computation-deadline  constraint was formulated with the optimization variable being  the input-data  transmission time, where the famous Shannon-Hartley formula gives the power-rate function. The optimization problem is convex  and can be solved in closed form. In particular, task offloading is desirable when the channel power gain is greater than a threshold and the server CPU is fast enough, which reveals the effects  of wireless channels on the offloading decision. A further study was conducted by Zhang \emph{et al.} in \cite{WZhangTWC1312} to minimize the energy consumption for executing a task with a soft real-time requirement, targeting e.g., multimedia applications, which requires the task to be completed within the deadline with a given  probability $\rho$. The offloading decision was determined by the computation mode (either offloading or local computing) that incurs less energy consumption. On one hand, the energy consumption for local execution was optimized using the DVFS technique, which was formulated as a convex optimization problem with the objective function being the expected energy consumption of the $W_{\rho}$ CPU cycles and a time duration constraint for these CPU cycles. The optimal CPU-cycle frequencies over the computation duration were derived in closed form by solving the \emph{Karush-–Kuhn-–Tucker} (KKT) conditions, suggesting that the processor should speed up as the number of completed CPU cycles increases. On the other hand, the expected energy consumption for task offloading was minimized via data transmission scheduling. Under the Gilbert-Elliott channel model, the optimal data transmission scheduling was obtained through \emph{dynamic programming} (DP) techniques, and the scaling law of the minimum expected energy consumption with respect to the execution deadline was also derived. This framework was further developed in \cite{you2016energyJSAC} where both the local computing and offloading are powered by wireless energy transfer. Specifically, the optimal CPU-cycle frequencies for local computing and time division for offloading should be adaptive to the transferred power.

\subsubsection{\textbf{Deterministic Task Model with Partial Offloading}}  The running of a relatively sophisticated mobile  application can be decomposed into a set of smaller sub-tasks. Inspired by recent advancements of parallel computing, partial offloading (also known  as program partitioning) schemes were proposed to further optimize MEC performance in \cite{YWangTCOM16,MJiaINFOCOMWS1404,MahmoodiTCC16,KaoINFOCOM1504,WZhangTWC1501,Khalili1508,PLorenzo1603,MahmoodiICC1506}.

In \cite{YWangTCOM16}, full granularity in program  partitioning  was considered where the task-input data can be arbitrarily divided for local and remote executions. Joint optimization of the offloading ratio, transmission  power and CPU-cycle frequency was performed  to minimize the mobile-energy consumption (or latency) subject to a latency (or energy consumption) constraint. Both the energy  and latency minimization problems are non-convex in contrast to the ones for binary-offloading.  The former problem can be  solved optimally with a variable-substitution technique while a sub-optimal algorithm was proposed for the latter one in \cite{YWangTCOM16}.

In \cite{MJiaINFOCOMWS1404,MahmoodiTCC16,KaoINFOCOM1504,WZhangTWC1501,Khalili1508,PLorenzo1603,MahmoodiICC1506}, applications were modeled by task-call graphs discussed earlier  that specify the dependency among different sub-tasks, and the code partitioning schemes   designed to dynamically generate the optimal set of tasks for offloading. In \cite{MJiaINFOCOMWS1404}, by leveraging the concept of load balancing between the mobile device and the server, a heuristic program-partitioning algorithm was developed to minimize the execution latency. Kao \emph{et al.} investigated the latency minimization problem with a prescribed resource utilization constraint in \cite{KaoINFOCOM1504}, and proposed a polynomial-time approximate solution with guaranteed performance. To maximize the energy savings achieved by computation offloading, the scheduling and cloud offloading decisions were jointly optimized using an integer programming approach in \cite{MahmoodiTCC16}. In \cite{WZhangTWC1501}, considering the wireless channel models including the block fading channel, \emph{independent and identical distributed} (i.i.d.) stochastic channel, and the Markovian stochastic channel, the expected energy consumption minimization problem with a completion time constraint was found to be a \emph{stochastic shortest-path} problem, and the \emph{one-climb} policies (i.e., the execution only migrates once from the mobile device to the server) were shown to be optimal. In addition, the program-partitioning schemes were also optimized together with the physical layer parameters, such as the transmission and reception power, constellation size, as well as the data allocation for different radio interfaces \cite{Khalili1508,PLorenzo1603,MahmoodiICC1506}.

\subsubsection{\textbf{Stochastic Task Model}} Resource management policies have been also developed for MEC systems with stochastic task models characterized by random task arrivals, where the arrived but not yet executed tasks join the queues in buffers \cite{DHuangTWC1206,JLiuISIT1607,SChenGLOBECOM1312,hong2016qoe,KwakJSAC1512,ZJiangAccess1511}. For such systems, the long-term performance, e.g., the long-term average energy consumption and execution latency, are more relevant compared with those of deterministic task arrivals, and the temporal correlation of the optimal system operations makes the design more challenging. As a result, the design of MEC systems with random task arrivals is an area less explored compared with the simpler cases with deterministic task models.  In \cite{DHuangTWC1206}, in order to minimize the mobile-energy consumption while keeping the proportion of executions violating the deadline  requirement below a threshold, a dynamic offloading algorithm was proposed to determine the offloaded software components from an application running at a mobile  user based on Lyapunov optimization techniques, where 3G and WiFi networks are accessible to the device but their rates vary at different locations. Assuming that concurrent local and edge executions are feasible, the latency-optimal task scheduling policies were designed  in \cite{JLiuISIT1607} based on the theory of \emph{Markov decision process} (MDP), which controls  the states of the local processing and transmission units and the task buffer queue length based on the  channel state. It was shown that the optimal task-scheduling policy significantly  outperforms the greedy scheduling policy (i.e., tasks are scheduled to the local CPU/transmission unit whenever they are idle).  To jointly optimize the computation  latency and energy consumption, the problem of minimizing the  long-term average execution cost  was considered in \cite{MahmoodiICC1506} and \cite{hong2016qoe}, where the former only optimized the offloading data size based on the  MDP theory while the latter jointly controlled the local CPU frequency, modulation scheme as well as data rates under a semi-MDP framework. In \cite{KwakJSAC1512}, the energy-latency tradeoff in MEC systems with heterogeneous types of applications was investigated, including the non-offloadable workload, cloud-offloadable workload and network traffic. A Lyapunov optimization-based algorithm was proposed to jointly decide the offloading policy, task allocation, CPU clock speed, and selected network interface. It was also shown that the energy consumption decreases inversely proportional to $V$ while the latency increases linearly with $V$, where $V$ is a control parameter in the proposed algorithm. Similar investigation was conducted for MEC systems with a multi-core mobile device in \cite{ZJiangAccess1511}.

{\color{black}{\subsubsection{Summary and Insight}
The comparison of resource management schemes for single-user MEC systems is shown in Table~\ref{Tab:Singleuser}. This series of work yields a number of useful insights on controlling computation offloading as summarized below.
\begin{itemize}
\item Consider binary offloading. For energy savings, computation offloading is preferred to local computation when the user has desirable channel condition or small local computation capability. Moreover, beamforming and MIMO techniques can be exploited to reduce the energy consumption for offloading. For latency reduction, computation offloading is advantageous over local computation when the user has a large bandwidth and the MEC server is provisioned with huge computation capacity.
\item Partial offloading allows flexible components/data partitioning. By offloading time-consuming or energy-consuming sub-tasks to MEC servers, partial offloading can achieve larger energy savings and smaller computation latency compared with binary offloading. Graph theory is a powerful tool for designing the offloading scheduling  according to the task dependency graph.
\item For stochastic task models, the temporal correlation of task arrivals and channels can be exploited to design adaptive dynamic computation offloading policies. Moreover, it is critical to maintain the task buffer stability at the user and MEC server via offloading rate control.
\end{itemize}

\begin{table*}\footnotesize
\centering
 \caption{\color{black}{The comparison of papers focusing on single-user MEC systems.}}
 \label{Tab:Singleuser}
 \color{black}{
\begin{tabular}{ | p{2.3cm}|  p{2.8cm}|  p{1.3cm}|  p{8cm}|}
 \hline
 \textbf{Task model} &  \textbf{Design Objective}  &  \textbf{Reference} &  \textbf{Proposed Solution}  \\
 \hline
 \multirow{8}{*}{Binary Offloading} & \multirow{6}{*}{Energy} 
 & \cite{WZhangTWC1312} & Optimize local computing and offloading by controlling the CPU frequency and transmission rate\\ \cline{3-4}
   &  & \cite{you2016energyJSAC} & Propose a novel framework of wirelessly powered MEC and optimize both local computing and offloading \\ \cline{3-4}
   & & \cite{KumarCompt1004} & Propose general guidelines to make offloading decision for energy consumption minimization \\ \cline{3-4}
  &  & \cite{BarbarossaMSP1411} & Propose the optimal binary computation offloading decision using convex optimization \\

 \cline{2-4}
   & \multirow{2}{*}{Energy and latency}  & \cite{KumarMPA1004} & Propose general guidelines to make offloading decision for energy-consumption and computation-latency minimization \\ \cline{3-4} 
 \hline
 \multirow{14}{*}{Partial Offloading} & \multirow{11}{*}{Energy} 
   & \cite{MahmoodiTCC16} & Propose a joint scheduling and computation offloading algorithm by parallel processing appropriate components in the mobile and cloud \\ \cline{3-4}
  &   & \cite{WZhangTWC1501} & Formulate a stochastic shortest-path problem and derive the one-climb optimal policy \\ \cline{3-4}
  &  & \cite{PLorenzo1603} & Jointly optimize the program partitioning with the selection of transmit power and constellation size \\ \cline{3-4}
  &  & \cite{MahmoodiICC1506} & Propose an iterative algorithm for the optimal offloading scheduling as well as the percentage of the data to be carried on each radio interface \\ \cline{2-4}
  &  \multirow{2}{*}{Latency}  & \cite{MJiaINFOCOMWS1404} & Propose a heuristic load-balancing program-partitioning algorithm \\ \cline{3-4}
 & &\cite{KaoINFOCOM1504} & Propose a polynomial-time approximate solution with guaranteed performance\\ \cline{2-4}
  & \multirow{3}{*}{Energy and latency}  & \cite{YWangTCOM16}  & Jointly optimize the offloading ratio, transmission power and CPU-cycle frequency using variable-substitution technique\\ \cline{3-4}
  &   & \cite{Khalili1508} & Propose an algorithmic to leverage the structure of the call graphs  by means of message passing  under both serial and parallel implementations of processing and communication \\ \cline{2-4}
 \hline
  \multirow{12}{*}{Stochastic Model} & \multirow{2}{*}{Energy} 
   & \cite{DHuangTWC1206} & Propose a Lyapunov optimization-based dynamic computation offloading policy \\ \cline{2-4}
   & \multirow{4}{*}{Latency}  & \cite{JLiuISIT1607} & Dynamically control the local processing and transmission using MDP \\ \cline{3-4} 
  &   & \cite{SChenGLOBECOM1312} & Optimize local computing and transmission using semi-MDP and propose a one-dimensional heuristic search algorithm \\ \cline{2-4}
 & \multirow{8}{*}{Energy and Latency}  & \cite{hong2016qoe} & Jointly control the local CPU frequency, modulation scheme as well as the data rates under a semi-MDP framework \\ \cline{3-4} 
   &    & \cite{KwakJSAC1512}  & Propose  a Lyapunov optimization-based algorithm  to  decide the offloading policy, task allocation, CPU clock speed, and selected network interface \\ \cline{3-4}
  &   & \cite{ZJiangAccess1511} & Propose a Lyapunov optimization-based scheme for cloud offloading scheduling, as well as download scheduling for cloud execution output \\ \cline{2-4}
 \hline
 \end{tabular}}
 \end{table*}

}}
\subsection{Multiuser MEC Systems}

While the preceding subsection aims at resource management policies for single-user MEC systems with a dedicated MEC server, this subsection considers the multiuser MEC systems comprising multiple mobile devices that share one edge server. Several new challenges are investigated in the sequel, including the multiuser joint radio-and-computational resource allocation, MEC server scheduling, and multiuser cooperative edge computing.

\subsubsection{\textbf{Joint Radio-and-Computational Resource Allocation}}
Compared with the central cloud, the MEC servers have much less  computational resources. Therefore, one key issue in designing  a multiuser MEC system is how to allocate the finite radio-and-computational  resources to multiple mobiles for achieving a system-level  objective, e.g., the minimum sum mobile-energy consumption. Both the centralized and distributed resource allocation schemes have been studied for different MEC systems as reviewed in the following..

For  centralized resource allocation
\cite{you2016energy,barbarossa2013joint,PLorenzo1603,hoang2012optimal,YMao1612GC,KWang16TCC,ren2017latency,MHChen1605,MHChen1704}, the MEC server obtains all the mobile information, including the CSI and computation requests, makes the resource-allocation decisions, and informs the mobile devices about the decisions. In \cite{you2016energy}, mobile users time-share a single edge server and have different computation workloads and local-computation capacities. A convex optimization problem was formulated to minimize the sum mobile-energy consumption. The key finding is that the optimal policy for controlling offloading data size and time allocation has a simple threshold-based structure. Specifically, an offloading priority function was firstly derived according to mobile users' channel conditions and local computing energy consumption. Then, the users with priorities above and below a given threshold will perform full and minimum offloading (so as to meet a given computation deadline), respectively. This result was also extended to the OFDMA-based MEC systems for designing a close-to-optimal computation offloading policy. In \cite{barbarossa2013joint}, instead of controlling the offloading data size and time, the MEC server determined the mobile-transmission  power and assigned server CPU cycles to different users in order to reduce the sum mobile-energy consumption. The optimal solution shows that, there exists an optimal  one-to-one mapping  between the transmission   power and the number of allocated CPU cycles for each mobile device. This work was further extended in \cite{PLorenzo1603} to account for the optimal binary offloading based on the model of task-call graphs. In \cite{ren2017latency}, the authors considered the multiuser video compression offloading in MEC and minimized the latency in local compression, edge cloud compression and partial compression offloading scenarios.
 Besides, in order to minimize the energy and delay cost for multi-user MEC systems where each user has multiple tasks, Chen \emph{et al.} jointly optimized the offloading decisions and the allocation of communication resource via a \emph{separable semidefinite relaxation} approach in \cite{MHChen1605}, which was later extended in \cite{MHChen1704} by taking the computational resource allocation and processing cost into account. Different from \cite{you2016energy,barbarossa2013joint,PLorenzo1603,MHChen1605,ren2017latency,MHChen1704}, the revenue of service providers was maximized in \cite{hoang2012optimal} under constraints of \emph{quality of service} (QoS) requirements for all mobile devices. The assumed fixed resource usage of each user results in a semi-MDP problem, which was transformed into a \emph{linear programming} (LP) model and efficiently solved. In \cite{YMao1612GC},  assuming a stochastic task arrival model, the energy-latency tradeoff in multiuser MEC systems was investigated via a Lyapunov optimization-based online algorithm, which jointly manages the available radio-and-computational resources. Centralized resource management for multiuser MEC system based on \emph{cloud radio access network} (C-RAN) has also been investigated in \cite{KWang16TCC}.

Another thrust  of research targets  distributed resource allocation for multiuser MEC systems which were designed  using game theory and decomposition techniques \cite{XChen1504,chen2015efficient,MHChen1610,ma2015game,lyumulti:2016:ProxiCloud,guo2016energy,Sardellitti1506}. In \cite{XChen1504} and \cite{chen2015efficient}, the computation tasks were assumed to be either locally executed or fully offloaded via single and multiple interference channels, respectively. With fixed mobile-transmission  power, an integer optimization problem was formulated to minimize the total energy consumption and offloading latency, which was proved to be NP-hard. Instead of designing a centralized solution, the game-theoretic techniques were applied to develop  a distributed algorithm that is able to achieve a Nash equilibrium. Moreover, it was shown that for each user, offloading is beneficial only when the received interference power is lower than a threshold. Furthermore, this work was extended in \cite{MHChen1610} and \cite{ma2015game}, where each mobile has multiple tasks and can offload computation to multiple APs
connected by a common edge-server, respectively. For the offloading process, in addition to transmission energy, this work has also accounted for the scanning energy of the APs and the fixed circuit power. The proposed distributed offloading policy showed that a mobile device should  handover the computation to a different  AP only when a new user choosing the same AP achieves a larger benefit. Building on the system model in \cite{chen2015efficient}, the joint optimization for the mobile-transmission power and the CPU-cycle allocation of the edge server was investigated in \cite{lyumulti:2016:ProxiCloud}. To solve the formulated mixed-integer problem, the decomposition technique was utilized to optimize the resource allocation and offloading decision sequentially. Specifically, the offloading decision problem was reduced to a sub-modular maximization problem and solved by designing a heuristic greedy algorithm. Similar decomposition technique and successive convex approximation technique were  utilized in \cite{guo2016energy} and \cite{Sardellitti1506} respectively to design distributed resource allocation algorithm for MEC systems.

\subsubsection{\textbf{MEC Server Scheduling}}
The  works discussed earlier  \cite{you2016energy,barbarossa2013joint,hoang2012optimal,chen2015efficient,ma2015game,lyumulti:2016:ProxiCloud} are based on the assumptions of user synchronization and the feasibility of parallel local-and-edge computation. However, studying practical  MEC server scheduling requires relaxation of these assumptions as discussed below together with  the resultant designs.
First, the arrival times of different users are in general asynchronous so that it is desirable for the edge server with finite computational resource to buffer and compute the tasks sequentially, which incurs the queuing delay. In \cite{molina2014joint}, to cope with the bursty task arrivals, the server scheduling was integrated with uplink-downlink transmission scheduling to minimize the average latency using  queuing theory. Second, even for synchronized task arrivals, the latency requirements can differ significantly over  users running  different types of applications ranging from latency-sensitive to  latency-tolerant applications. This fact calls for the server  scheduling assigning users different levels of priorities based on their latency requirements.  In \cite{yu2016joint}, after the pre-resource allocation, the MEC server will check the deadline of different tasks during the server computing process and adaptively adjust the task execution order to satisfy the heterogeneous  latency requirements. Last, some computation tasks each consists of  several dependent sub-tasks such that the scheduling of these modules must satisfy the task-dependency requirements.  The task model with a sequential sub-task arrangement   was considered in\cite{yang2015multi} that  jointly optimizes the program partitioning for multiple users and the server-computation scheduling to minimize the average completion time. As a result, a heuristic algorithm was proposed to solve the formulated mixed-integer problem. Specifically, it first optimizes the computation partition for each user. Under these partitions, it will search the time intervals violating the resource constraint and adjust them accordingly. Furthermore, the general dependency-task model as shown in Fig.~\ref{CG:gen} was considered for multiple users in \cite{guo2016energy}. This model drastically complicates the computing time characterization. To address this challenge, a measure of \emph{ready time} was defined for each sub-task  as the earliest time when all the predecessors have been computed. Then, the offloading decision, mobile CPU-cycle frequency and mobile-transmission power were jointly optimized to reduce the sum mobile-energy consumption and computation latencies with a proposed distributed algorithm.
\begin{table*}[h!]\footnotesize
\centering
  \caption{\color{black}{The comparison of papers focusing on multiuser MEC systems.}}
  \label{Tab:Multiuser}
  \color{black}{
\begin{tabular}{ | p{2.1cm}|  p{2.1cm}|  p{2.5cm}| p{1.6cm}|  p{7cm}|}
 \hline
 \textbf{Theme} &  \textbf{Design Type/Motivation} &   \textbf{Design Objective} &  \textbf{Reference} &  \textbf{Proposed Solution}  \\
 \hline
 \multirow{29}{*}{} & \multirow{14}{*}{Centralized} 
  &  \multirow{9}{*}{Energy} & \cite{you2016energy} & Design the optimal threshold-based resource allocation policy  based on defined offloading priority function for TDMA and OFDMA systems \\ \cline{4-5}
   & & & \cite{barbarossa2013joint} & Jointly optimize the allocation of communication and computation resources \\ \cline{4-5}
     & & & \cite{PLorenzo1603} & Design the optimal resource allocation and code partitioning by call-graph selection approach  \\ \cline{4-5}
     & & & \cite{KWang16TCC} & Solve the non-convex resource allocation problem for C-RAN using iterative algorithms \\ \cline{3-5}
     &  & \multirow{2}{*}{Latency} & \cite{ren2017latency} & Minimize the latency in multiuser video compression via resource allocation \\ \cline{3-5}
   &  & \multirow{4}{*}{Energy and latency} & \cite{YMao1612GC} & Propose a Lyapunov optimization-based dynamic computation offloading policy \\ \cline{4-5}
 & & & \cite{MHChen1605,MHChen1704} & Jointly optimize the offloading decisions and the allocation of resource via semidefinite relaxation\\
 \cline{3-5}
  Joint radio-and-computational&  & Revenue & \cite{hoang2012optimal} & Design the optimal resource allocation based on semi-MDP\\
 \cline{2-5}
   resource allocation& \multirow{11}{*}{Distributed}  &  \multirow{2}{*}{Energy} & \cite{Sardellitti1506} & Propose a distributed iterative algorithm using successive convex approximation technique \\ \cline{3-5} 
  & & \multirow{9}{*}{Energy and latency}  &   \cite{XChen1504,chen2015efficient} & Develop a distributed algorithm that is able to achieve a Nash equilibrium \\ \cline{4-5}
    & & & \cite{MHChen1610} & Propose a distributed algorithm for multi-user MEC systems where each user has multiple tasks \\ \cline{4-5}
  & & & \cite{ma2015game} & Consider multiple servers and develop a distributed algorithm admitting the Nash equilibrium \\ \cline{4-5}
  & & &  \cite{guo2016energy } & Propose a decomposition algorithm  to control the computation offloading selection, clock frequency control and transmission power allocation iteratively \\ \cline{3-5}
  &  & \multirow{2}{*}{Utility} & \cite{lyumulti:2016:ProxiCloud} & Propose a decomposition algorithm  to optimize the resource allocation and offloading decisions\\  \cline{3-5}

 \hline
 \multirow{7}{*}{MEC server} & Bursty data arrivals 
  &  \multirow{2}{*}{Latency} & \cite{molina2014joint} & Optimize the uplink and downlink scheduling using queuing theory \\ \cline{2-5}
  & Heterogeneous deadlines & \multirow{2}{*}{Energy} & \cite{yu2016joint} & Propose a pre-resource allocation and joint scheduling scheme \\ \cline{2-5}
 scheduling & \multirow{4}{*}{Task dependency} &\multirow{2}{*}{Latency} & \cite{yang2015multi} &  Propose heuristic algorithm with searching and adjusting phases based on constraint relaxation \\ \cline{3-5}
  & & \multirow{2}{*}{Energy and latency}& \cite{guo2016energy} & Propose a decomposition algorithm  to control the computation offloading selection, clock frequency control and transmission power allocation iteratively \\ \cline{1-5}

  \multirow{13}{*}{Cooperative} & \multirow{1}{*}{D2D } 
  &  \multirow{1}{*}{Task success rate} & \cite{li2014exploring} & Propose the optimal and periodic mobile cloud access scheme \\ \cline{3-5}
  &communication & Network capacity and offloading probability & \cite{jo2015device} & Propose D2D communication techniques in heterogeneous MEC systems \\ \cline{2-5}
 & \multirow{4}{*}{Cooperation}  & \multirow{4}{*}{Energy} &  \cite{sheng:2016:energy} & Propose a fairness-aware energy-efficient cooperative node selection scheme \\ \cline{4-5}
    & & & \cite{XCao1704} & Propose a four-slot protocol to enable joint computation and communication cooperation \\ \cline{2-5}
 computing &  Share computation results & \multirow{2}{*}{Energy} & \cite{song2014energy} & Propose a Lyapunov optimization-based cooperative computing policy \\ \cline{2-5}
  & Share computational resource & \multirow{2}{*}{Energy} & \cite{You:2017aa} & Propose a ``string-pulling" offloading policy based on constructed offloading feasibility tunnel \\ \cline{2-5}
    & Small BSs cooperation & \multirow{2}{*}{Delay cost} & \cite{LChen1703} & Propose a peer offloading framework that allows both centralized and autonomous decision making \\ \cline{2-5}
 \hline
 \end{tabular}}
 \end{table*}

\subsubsection{\textbf{Multiuser Cooperative Edge Computing}}
Multiuser cooperative computing is envisioned as a promising technique to improve the MEC performance by providing  two advantages \cite{li2014exploring,jo2015device,sheng:2016:energy,song2014energy,XCao1704,You:2017aa,LChen1703}. First, MEC servers with limited computational resources may  be overloaded when they have to serve  a large number of offloading mobile users. In such cases, the  burdens on the servers  can be lightened via peer-to-peer mobile  cooperative computing. Second, sharing the computational resources among the users can balance the uneven distribution  of the  computation workloads and computation capabilities over  users.
In \cite{li2014exploring}, D2D communication was proposed to enable multiuser cooperative computing. In particular, this work studied  how to detect and utilize computational resources on other users. This idea was  adopted in \cite{jo2015device} to propose a D2D-based heterogeneous MCC networks. Such a novel framework was shown to enhance the  network capacity and offloading probability. Moreover, for wireless sensor networks, cooperative computing was proposed in \cite{sheng:2016:energy} to enhance its computation capability.  First, the optimal computation partition for minimizing the total energy consumption of two cooperative nodes was investigated. This result was then utilized to design the fairness-aware energy-efficient cooperative node selection. Furthermore, Song \emph{et al.} showed that sharing computation results among the peer users can significantly reduce the communication traffic for a multiuser MEC system \cite{song2014energy}. Assuming the task can either be offloaded or computed locally, a mixed-integer optimization problem was formulated to minimize the total energy consumption under the constraint of the system communication traffic. To tackle this challenging problem, two online task scheduling algorithms were proposed based on pricing and Lyapunov optimization theories.  In addition, by employing a helper, a four-slot joint computation-and-communication cooperation protocol was proposed in \cite{XCao1704}, where the helper not only computes part of the tasks offloaded from the user, but also acts as a relay node to forward the tasks to the MEC server. Another recent work \cite{You:2017aa} investigated the optimal offloading policies in a peer-to-peer cooperative computing system where the computing helper has time-varying computation resources. Specifically, an \emph{offloading feasibility tunnel} was constructed based on the helper's CPU profile and buffer size. Given the tunnel, the optimal offloading was shown to be achieved by the well-known ``string-pulling" strategy, graphically referring to pulling a string across the tunnel. Last, Chen \emph{et al.} proposed an online peer offloading framework based on Lyapunov optimization and game theoretic approaches in \cite{LChen1703}, which enables small BSs cooperation to handle the spatially uneven computation workloads in the network. 

{\color{black}{\subsubsection{Summary and Insight}
The comparison of resource management schemes for multiuser MEC systems is provided in  Table~\ref{Tab:Multiuser}. We draw several  conclusions on resource allocation, MEC server scheduling and mobile cooperative computing as follows.
\begin{itemize}
\item Consider multiuser MEC systems with finite radio-and-computational resources. For system-lever objectives, e.g., to minimize the sum mobile energy consumption, the users with large channel gains and low local-computation energy consumption have higher priorities for offloading computation since they can contribute to larger energy savings. Too many offloading users, however, will cause severe inter-user interference of communication and computation, which will, in turn, reduce the system revenue.
\item To effectively reduce the sum computation latency of multiple users, the scheduling design for a MEC server should assign higher priorities to the users with more stringent latency requirements and heavy computation loads. Moreover, parallel computing can further boost the computation speed at the server.
\item Scavenging the enormous amount of distributed computation resources can not only alleviate the network congestion, but also improves resource utilization and enables ubiquitous computing. This vision can be materialized by peer-to-peer mobile cooperative edge computing. The key advantages include short-range transmission via D2D techniques and computation resource and result sharing.
\end{itemize}
}}
\subsection{MEC Systems with Heterogeneous  Servers}
To enable ubiquitous edge computing, \emph{heterogeneous MEC} (Het-MEC)  systems were proposed in \cite{lei2013challenges} comprising one central cloud and multiple edge servers. The coordination and interaction of multi-level central/edge clouds introduce many interesting new research challenges and recently have attracted extensive relevant investigations on  server selection, cooperation and computation migration, as discussed in the sequel.

\subsubsection{\textbf{Server Selection}} For users served by a Het-MEC system, a  key design  issue is to determine the destination of computation offloading, i.e., either the edge or central cloud server. In \cite{zhao2015cooperative}, the server selection problem was studied for a multiuser system comprising a single edge server and a single  central cloud. To maximize the total successful offloading probability, a heuristic scheduling algorithm was proposed to leverage both the low communication latency due to  the proximity of the  MEC server and the low computation latency arising from  abundant computational resources at the central-cloud server. Specifically, when the computation load of the MEC server exceeds a given threshold, latency-tolerant tasks are offloaded to the central cloud to spare enough computational resources at the edge server for processing latency-sensitive tasks. In addition, \cite{ge2012game} explored the problem of server selection over multiple MEC servers. The major challenge arises from the correlation between the amounts of the offloaded computation and selected edge servers for multiple users. To cope with this issue, a congestion game was formulated and solved to minimize the sum energy consumption of mobile users and edge servers. Most recently, a computation offloading framework that allows a mobile device to offload tasks to multiple MEC servers was proposed in \cite{TQDinh1704}, and semidefinite relaxation-based algorithms were proposed to determine the task allocation decisions and CPU frequency scaling.

\subsubsection{\textbf{Server Cooperation}} Resource sharing via server cooperation can not only improve the resource utilization and increase the revenue of computing service providers, but also provide more resources for mobile users to enhance their user experience. This framework was originally proposed in  \cite{kaewpuang2013framework}, which includes components such as resource allocation, revenue management and service provider cooperation. First, resource allocation was optimized for cases with deterministic and random user information to maximize the total revenues. Second, considering self-interested cloud service providers, a distributed algorithm based on game theory was proposed to maximize service providers' own profits, which was shown to achieve the Nash equilibrium. This study was further extended in \cite{yu2015decentralized}, which considered both the local and remote resource sharing. The former refers to resource sharing among different service providers within the same data center, while the latter one means the cooperation across different data centers. To realize the resource sharing and cooperation among different servers, a coalition game was formulated and solved by a game-theoretic algorithm with stability and convergence guarantees. Moreover, the recent work \cite{elbamby2017proactive} proposed a new server cooperation scheme where edge servers exploit both the computational and storage resources by proactively caching computation results to minimize the computation latency. The corresponding task distributing problem was formulated as a matching game  and solved by an efficient algorithm based on a proposed deferred-acceptance algorithm.

\begin{table*}[t!]\footnotesize
\centering
 \caption{ \color{black}{The comparison of papers focusing on MEC systems with heterogeneous servers.}}
  \label{Tab:Multiserver}
  \color{black}{
\begin{tabular}{ | p{2.3cm}|  p{2.1cm}|  p{2.3cm}| p{1.3cm}|  p{7cm}|}
 \hline
 \textbf{Theme} &  \textbf{Design Type} &   \textbf{Design Objective} &  \textbf{Reference} &  \textbf{Proposed Solution}  \\
 \hline

 \multirow{7}{*}{Server selection} & Edge/central server selection
  & Successful offloading probability &  \cite{zhao2015cooperative} & Propose a heuristic server selection algorithm according to the deadline requirements \\ \cline{2-5}
  &  Edge server selection 
  & Energy &  \cite{ge2012game} & Formulate a congestion game and propose a distributed algorithm admitting the Nash equilibrium \\ \cline{2-5}
 &Multiple edge servers &\multirow{2}{*}{Energy and latency} &  \cite{TQDinh1704} &  Propose semidefinite relaxation-based algorithms for task allocation decisions and frequency scaling  \\ \hline

   \multirow{5}{*}{Server cooperation}   & Edge server cooperation & \multirow{2}{*}{Revenue}&  \cite{kaewpuang2013framework} & Propose a distributed resource allocation algorithm admitting the Nash equilibrium  \\ \cline{2-5}
      &Edge/remote server cooperation & \multirow{2}{*}{Utility} & \cite{yu2015decentralized} & Formulate a coalition game and propose a game-theoretic algorithm \\ \cline{2-5}
         &Edge server proactive caching & \multirow{2}{*}{Latency} & \cite{elbamby2017proactive} & Study the distribution and proactive caching of computing tasks in MEC  \\ \hline

  \multirow{3}{*}{Computation}   & \multirow{3}{*}{Edge server } & \multirow{4}{*}{Cost}&  \cite{wang2014mobility} & Propose a threshold-based computation migration scheme according to the distance  \\ \cline{4-5}
  migration    & migration &  & \cite{urgaonkar2015dynamic} & Propose online workload scheduling and migration algorithms using Lyapunov optimization techniques \\ \cline{2-5}
  & Remote server migration & \multirow{2}{*}{Energy and latency} & \cite{chen2016joint} & Propose a heuristic two-stage algorithm including migration decision and resource allocation \\ \hline
 \end{tabular}}
 \end{table*}
\subsubsection{\textbf{Computation Migration}} In \cite{wang2014mobility,urgaonkar2015dynamic,chen2016joint}, apart from optimizing the offloading decisions, the authors also investigated the computation migration among different remote servers. Specifically, the computation migration over MEC servers was motivated by the mobility of offloading users. When a user moves closer to a new MEC server, the network controller can choose to migrate the computation to this server, or compute the task in the original server and then forward the results back to the user via the new server. The computation migration problem was formulated as an MDP problem based on a random-walk mobility model in \cite{wang2014mobility}. It was shown that the optimal policy has a threshold-based structure, i.e., the migration should be selected only when the distance of two servers is bounded by two given thresholds. This work was further extended in \cite{urgaonkar2015dynamic} where the workload scheduling in edge servers was integrated with the service migration to minimize the average overall transmission and reconfiguration costs using Lyapunov optimization techniques. Another computation migration framework was proposed in \cite{chen2016joint}, where the MEC server can either process offloaded computation tasks locally or migrate them to the central cloud server. An optimization problem was formulated to minimize the sum mobile-energy consumption and computation latency. This problem was solved by a heuristic two-stage algorithm, which first determines the offloading decision for each user by the semi-definite relaxation and randomization techniques, and then performs the resource allocation optimization for all the users.

{\color{black}{\subsubsection{Summary and Insight}
Table~\ref{Tab:Multiserver} provides the summary of resource management schemes for MEC systems with heterogeneous servers. The literature provides a set of insights on server selection, cooperation, and computation migration, described as follows.
\begin{itemize}
\item Consider MEC systems with multiple computation tasks and heterogeneous servers. To reduce the sum computation latency, it is desirable to offload latency-insensitive but computation-intensive tasks to remote central cloud server and latency-sensitive ones to the edge servers.
\item Server cooperation can significantly improve the computation efficiency and resource utilization at MEC servers. More importantly, it can balance the computation load distribution over the networks so as to reduce sum computation latency while the resources are better utilized. Moreover, the server cooperation design should consider temporal-and-spatial computation task arrivals and server's computation capacities, time-varying channels, and servers' individual revenue.
\item Computation migration is an effective approach for mobility management in MEC. The decision of migrate-or-not depends on the migration overhead, distances between users and servers, channel conditions, and servers' computation capacities. Specifically, when a user moves far away from its original MEC server, it is preferred to migrate the computation to nearby servers.
\end{itemize}
}

\subsection{Challenges}

In the preceding subsections, we have conducted a comprehensive survey on the state-of-the-art resource management techniques for MEC systems. However, the progress is still in the infancy stage and many critical factors have been overlooked for simplicity, which need to be addressed in future research efforts. In the following, we identify three critical research challenges for resource management in MEC that remain to be solved.

\subsubsection{\textbf{Two-Timescale Resource Management}} In most existing works, e.g., \cite{BarbarossaMSP1411,chen2015efficient,Munoz1510,Sardellitti1506,lyumulti:2016:ProxiCloud,yu2016joint}, wireless channels were assumed to remain static during the whole task execution process for simplicity. Nevertheless, this assumption may be unreasonable when the channel coherence time is much shorter than the latency requirement. For instance, at a carrier frequency of 2GHz, the channel coherence time can be as small as 2.5ms when the speed is 100km\slash h. For some mobile applications such as the MMORPG game PlaneShift\footnote{\url{http://www.planeshift.it/}}, the acceptable response time is 440ms and the excellent latency is 120ms\cite{SWang0912}. In such scenarios, the task offloading process may be across multiple channel blocks, necessitating the two-timescale resource management  for MEC. This problem is very challenging even for a single-user MEC system with deterministic task arrivals \cite{WZhangTWC1312}.

\subsubsection{\textbf{Online Task Partitioning}} For ease of optimization, existing literature tackling the task partitioning problems ignores the fluctuation of the wireless channels, and obtain the task partitioning decision before the start of the execution process. With such an offline task partitioning decision, the change of the channel condition may lead to inefficient or even infeasible offloading, which shall severely degrade the computation performance. To develop online task partitioning policies, one should incorporate the channel statistics into the formulated task partitioning problem, which may easily belong to an NP-hard problem even under a static channel. In \cite{WZhangTWC1501} and \cite{SWangIEEEAcess1702}, approximate online task partitioning algorithms were derived for applications with serial and tree-topology task-call graphs, respective, while solutions for general task models remain unexploited.

\begin{figure*}[!t]
\begin{center}
   \includegraphics[width=0.75\textwidth]{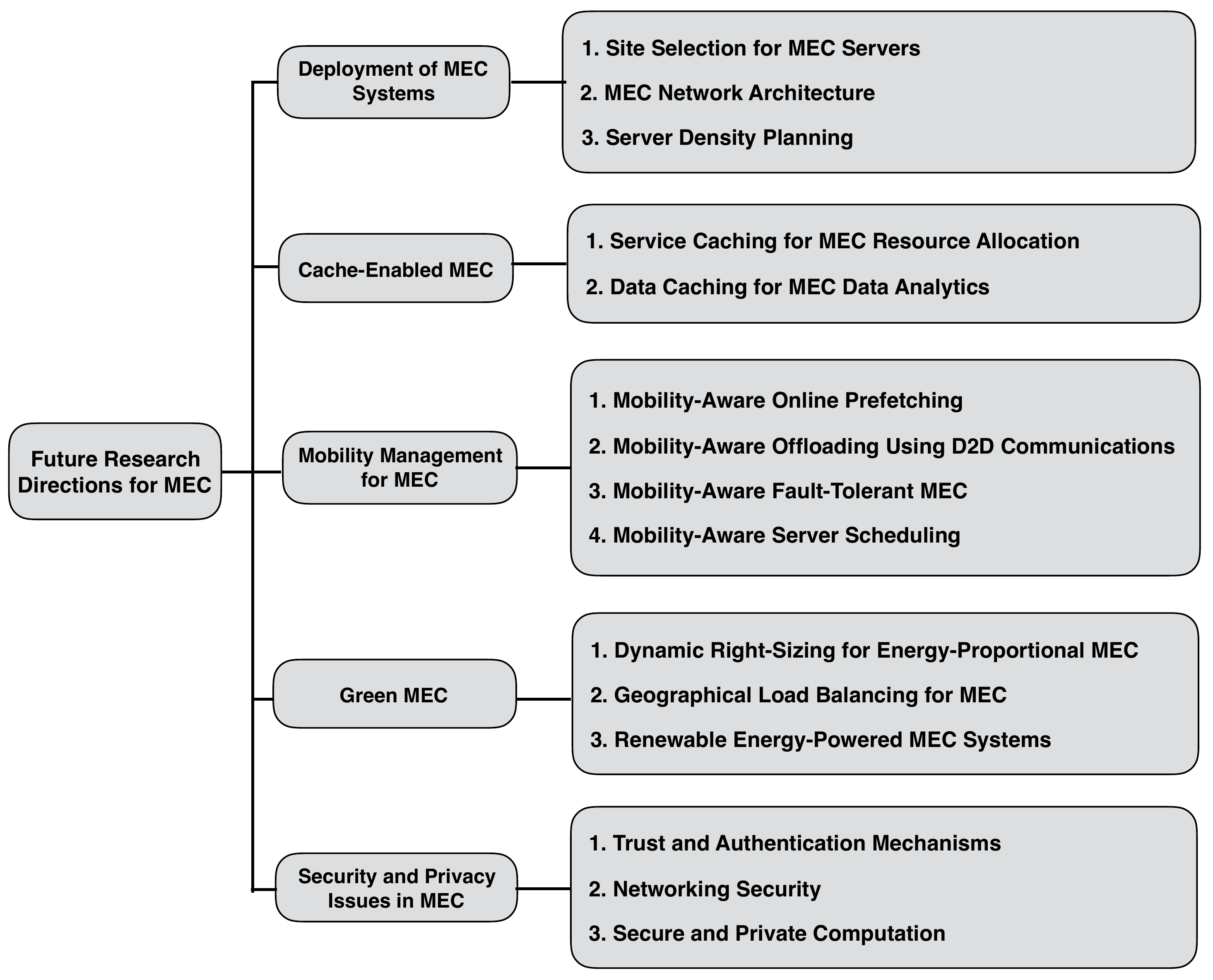}
\end{center}
\caption{Future research directions for MEC.}
\label{OutlookTree}
\end{figure*}
\subsubsection{\textbf{Large-Scale Optimization}} The collaboration of multiple MEC servers allows their resources to be jointly managed for serving a large number of mobile devices simultaneously. However, the increase of the network size renders the resource management a large-scale optimization problem with respect to a large number of offloading decisions as well as radio-and-computational resource allocation variables. {\color{black}Conventional centralized joint radio-and-computational resource management algorithms require a huge amount of information and computation when applied to large-scale MEC systems, which will inevitably incur a significant execution delay and may whittle away the potential performance improvement, e.g., latency reduction, brought by the MEC paradigm. To achieve efficient resource management, it is required  to design distributed low-complexity large-scale optimization algorithms with light signaling and computation overhead.} Although the recent advancements in large-scale convex optimization \cite{YShi1509} provide powerful tools for radio resource management, they cannot be directly applied to optimize the computation offloading decision due to its combinatorial and non-convex nature, which calls for new algorithmic techniques.

{\color{black}{\section{Issues, Challenges, and Future Research Directions}}}\label{Section:Others}

Recent years have witnessed substantial research efforts on resource management for MEC as surveyed in the preceding section. However, there are lots of emerging research directions of MEC  that are still largely uncharted. In this section, technical issues, challenges and research opportunities will be identified and discussed as summarized in Fig.~\ref{OutlookTree}, including the large-scale MEC system deployment, cache-enabled MEC, mobility management, green MEC and security-and-privacy issues in MEC.

\subsection{Deployment of MEC Systems}

The primary motivation of MEC is to shift the Cloud Computing capability to the network edges in order to reduce the latency caused by congestion and propagation delays in the core network. However, there is no formal definition of what an MEC server should be, and the server locations in the system are not specified. These invoke the site selection problems for MEC servers, which are significantly different from the conventional BS site selection problems, as the optimal placement of edge servers is coupled with the computational resource provisioning, and both of them are constrained by the deployment budget. Besides, the efficiency of an MEC system relies heavily on its architecture, which should account for various aspects such as workload intensity and communication rate statistics. In addition, it is critical for MEC vendors to determine the required server density for catering the service demand, which is closely related to the infrastructure deployment cost and marketing strategies. Nonetheless, the large-scale nature of MEC systems makes traditional simulation-based methods inapplicable, and thus solutions based on network-scale analysis are preferred. In this subsection, we will discuss three research problems related to MEC deployment, including the site selection for MEC servers, the MEC network architecture, and server density planning.

\subsubsection{\textbf{Site Selection for MEC Servers}}

Selecting the sites for MEC infrastructures, especially MEC servers, is the first step towards building up the MEC system. To make the cost-effective server-site selection, the system planners and administrators should account for two important factors: site rentals and computation demands. In general, given the system deployment budget, more MEC servers should be installed at regions with higher computation demands, such as business districts, commercial areas and densely populated areas. This, however, contradicts the cost requirement as such areas are likely to have high site rentals. Fortunately, thanks to the well-deployed telecom networks, it is a promising idea to install the MEC servers co-located with the existing infrastructures such as macro BSs, which is even more attractive for the telecom operators who would like to participate in the MEC market.

However, this would not solve all the problems. On one hand, due to the ever-increasing computation-quality requirement and ubiquitous smart devices, satisfactory user experience cannot be guaranteed due to the poor signal quality and congestion in the macro cells. For some applications, e.g., smart home \cite{CVallati1610}, it is desirable to move the computation capability even closer to the end users. This can be achieved by injecting some computational resources at small-cell BSs \cite{Andrews1203,Dhillon1203}, which are low-cost and small-size BSs. Despite the potential benefits, there are still obstacles on the way:

\begin{itemize}
\item First, due to physical limitations, the computation capabilities of such kind of MEC servers will be much smaller than those at  macro BSs, making it challenging  to handle  computation-intensive tasks. One feasible solution is to build a hierarchical network architecture for MEC systems comprising MEC servers with heterogeneous communication-and-computation capabilities as detailed in the sequel.

\item Second, some of the small-cell BSs may be self-deployed by the home users, and many femto BS owners may not have the motivation to collaborate with MEC vendors. To overcome this issue, MEC vendors need to design a proper incentive mechanism in order to stimulate the owners of small-cell BSs for renting the sites.

\item Moreover, deploying MEC servers at small-cell BSs may incur security problems as they are easy-to-reach and vulnerable to external attacks, which shall degrade the levels of reliability.
\end{itemize}

On the other hand, the computation hot spots do not always coincide with the communication hot spots. In other words, for some of the computation hot spots, there exists no available communication infrastructure (either macro or small-cell BS). For these circumstances, we need to deploy edge servers with wireless transceivers by properly choosing new locations.

Besides, the site selection for MEC servers is dependent on the computational resource-allocation strategy, which poses extra challenges compared to the conventional BS site selection. Intuitively, concentrating the computational resources at a few MEC servers can help save the site rentals. However, this comes at the prices of potential degradation of the service coverage and communication quality. In addition, the optimal computational resource allocation should take into account both site rentals and computation demands. For example,  for an MEC server at a site with a high site rental, it is preferred to allocate huge computational resource and thus serve a large number of users, for achieving the high revenue.  Hence, a joint site selection and computational resource provisioning problem needs to be solved before deploying MEC systems.

\subsubsection{\textbf{MEC Network Architecture}}\label{Sec:NetArchi}
\begin{figure}[!t]
\begin{center}
   \includegraphics[width=0.5\textwidth]{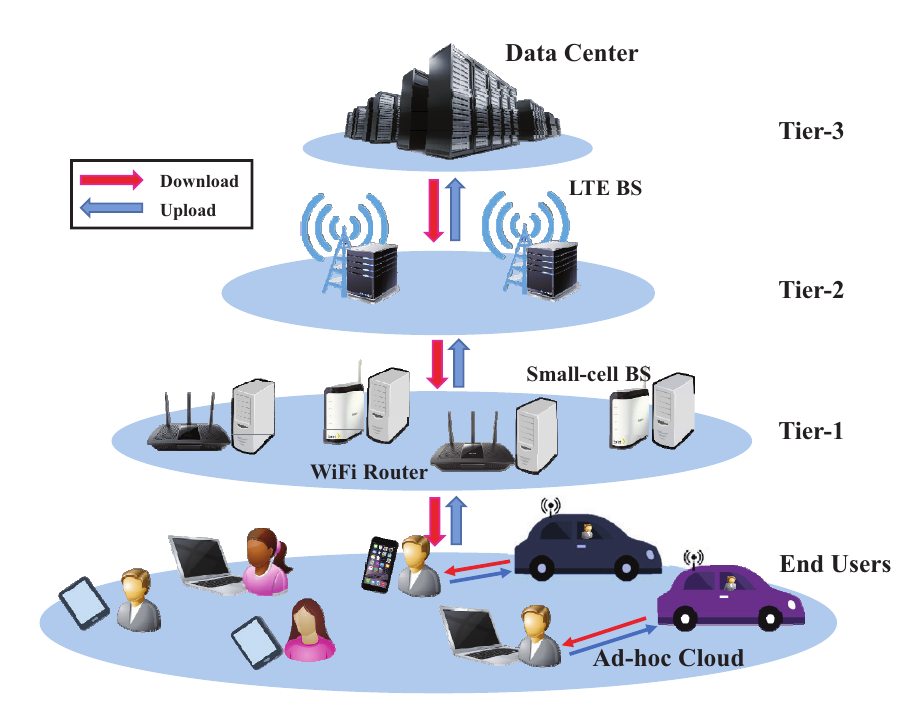}
\end{center}
\caption{A 3-tier heterogeneous MEC system. Tier-1 servers are located in close proximity to the end users, such as at  WiFi routers and small-cell BSs, which are of relatively small computation capabilities. Tier-2 servers are deployed at LTE BSs with moderate computation capabilities. Tier-3 servers are the existing Cloud Computing infrastructures, such as data centers.}
\label{HetCloud}
\end{figure}
The promotion of MEC does not mean the extinction of the \emph{data-center networks} (DCNs). Instead, future mobile computing networks are envisioned to  be consisted of three layers as shown in Fig.~\ref{HetCloud}, i.e., cloud, edge (a.k.a. fog layer), and the service subscriber layer \cite{lei2013challenges,TLuan1603}. While the cloud layer is mature and well-deployed, there is still some flexibility and uncertainty in designing the edge layer.

By analogy to the \emph{heterogeneous networks} (HetNets) in cellular systems, it is intuitive to design the Het-MEC systems, which consist of multiple tiers. Specifically, the MEC servers in different tiers have distinct computation and communication capabilities. Such kinds of hierarchical MEC system structures can not only preserve the advantage of efficient transmission offered by HetNets, but also possess strong ability to handle the peak computation workloads by distributing them across different tiers \cite{LTongINFOCOM1604}. However, the computation capacity provisioning problem is highly challenging and remains unsolved, as it should account for many different factors, such as the workload intensity, communication cost between different tiers, workload distribution strategies, etc.

Another thrust of research efforts focuses on exploiting the potential of the service subscriber layer, and utilizing the undedicated computational resources, e.g., laptops, smart phones, and vehicles, overlaid with dedicated edge nodes. This paradigm is termed as the \emph{Ad-hoc mobile cloud} in literature \cite{Kirby10,TTHuuCloudCom1412,ShilaWC16,XHouTVT1606}. The ad-hoc mobile cloud enjoys the benefits of amortizing the stress of MEC systems, increasing the utilization of the computational resources, and reducing the deployment cost. However, it also brings difficulties in resource management and security issues due to its ad-hoc and self-organized nature.

\subsubsection{\textbf{Server Density Planning}}
\begin{figure}[!t]
\begin{center}
   \includegraphics[width=0.5\textwidth]{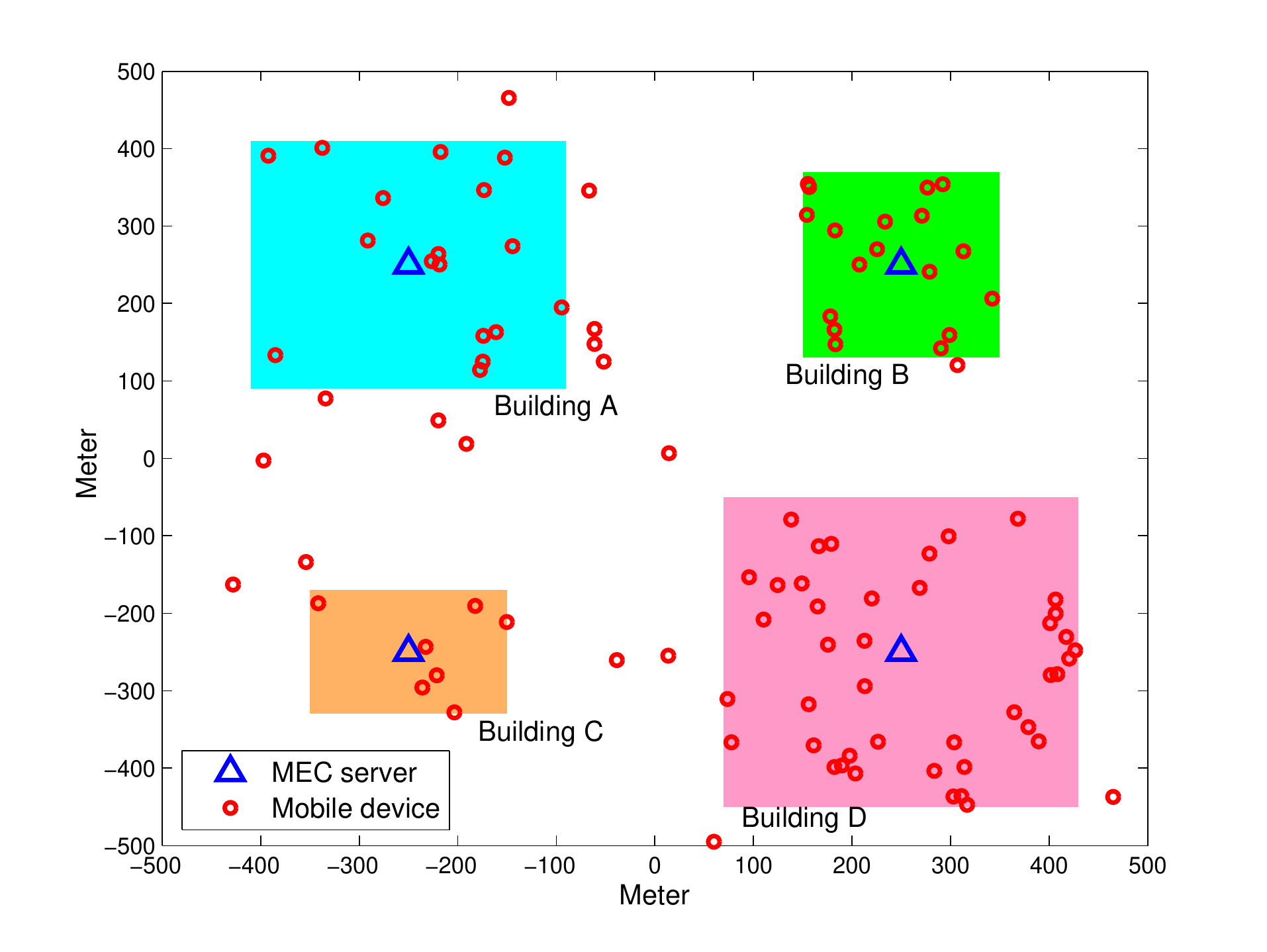}
\end{center}
\caption{Illustration of the clustering behavior of the computation demands. The mobile devices requesting for MEC services will be more concentrated around the MEC servers.}
\label{ClusterBehavior}
\end{figure}
As mentioned in Section~\ref{Sec:NetArchi}, the MEC infrastructure may be a combination of different types of edge servers, which provides various levels of computation experience and contributes different deployment costs. Hence, it is critical to determine the number of edge nodes as well as the optimal combination of different types of MEC servers with a given deployment budget and computation demand statistics. Conventionally, this problem can only be addressed by numerical simulations, which is time-consuming and has poor scalability. Fortunately, owing to the recent development of \emph{stochastic geometry theory} and its successful applications in performance analysis for wireless networks \cite{Haenggi0909,JGAndrews1111,Haenggi2012,CLiTCOM1604}, as well as the similarity between Het-MEC systems and HetNets, it is feasible to conduct performance analysis for MEC systems using techniques from stochastic geometry theory. Such analysis of MEC systems should address the following challenges: 1) The timescales of computation and wireless channel coherence time may be different \cite{WZhangTWC1312,JLiuISIT1607}, which makes existing results for wireless networks not readily applicable for MEC systems. One possible solution is to combine the Markov chain and stochastic geometry theories to capture the steady behavior of computations. 2) The computation offloading policy will affect the radio resource management policy, which should be taken into consideration. 3) The computation demands are normally non-uniformly distributed and clustered (see Fig. \ref{ClusterBehavior}), prohibiting the use of the \emph{homogeneous Poisson point process} (HPPP) model for edge servers and service subscribers. It thus calls for the investigation of more advanced point processes, e.g., the Ginibre $\alpha$-\emph{determinantal point process} (DPP), to capture the clustering behaviors of edge nodes \cite{VastardisDPP1402}.

\subsection{Cache-Enabled MEC}

It has been predicted by Cisco that mobile video streaming will occupy up to 72\% of the entire mobile data traffic by 2019 \cite{index2013global}. One unique property of such services is that the content requests are highly concentrated and some popular contents will be asynchronously and repeatedly requested. Motivated by this fact, \emph{wireless content caching} or \emph{FemtoCaching} was proposed in \cite{wang2014cache,bastug2014living,golrezaei2012femtocaching,gomes2017edge} to avoid frequent replication for the same contents by caching them at BSs. This technology has attracted extensive attention from both academia and industry due to its striking advantages on reducing content acquisition latency, as well as relieving heavy overhead burden of the network backhaul. While caching is to move popular contents close to end users, MEC is to deploy edge servers to handle computation-intensive tasks for edge users to enhance user experience. Note that these two techniques seem to target for diverse research directions, i.e., one for popular content delivery and the other for individual computation offloading. However, they will be integrated seamlessly in this subsection and envisioned to create a new research area, namely, the \emph{cache-enabled MEC}.

\begin{figure}[!t]
\begin{center}
   \includegraphics[width=0.4\textwidth]{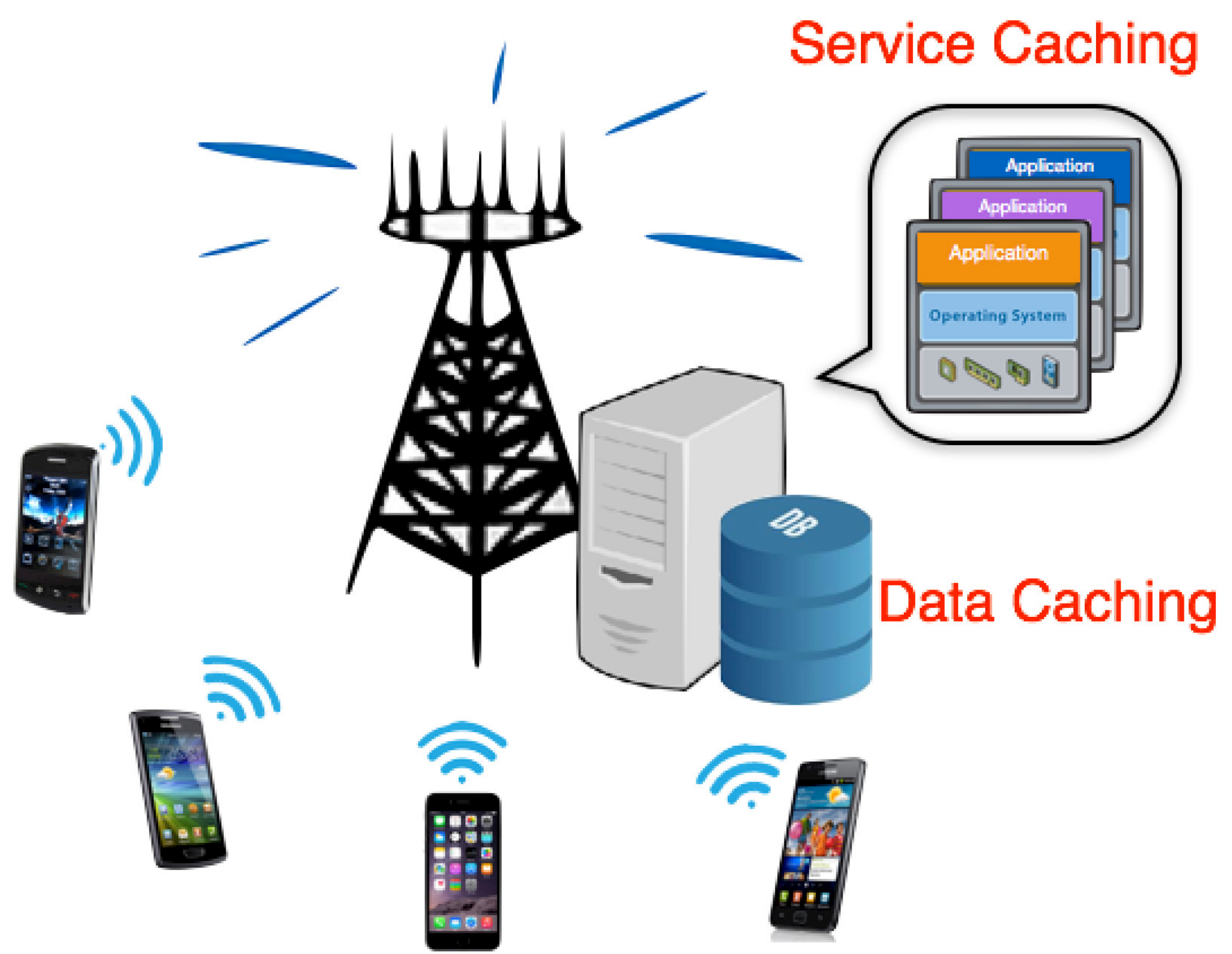}
\end{center}
\caption{Cache-enabled MEC systems.}
\label{FigCacheMEC}
\end{figure}
Consider the novel cache-enabled MEC system shown in Fig.~\ref{FigCacheMEC}. In such systems, the MEC server can cache several application services and their related database, called \emph{service caching} (or service placement \cite{yang2016cost}) and \emph{data caching}, respectively, and handle the offloaded computation from multiple users. To efficiently reduce the computation latency, several key and interesting problems need to be solved, which are described in the following with potential solutions.

\subsubsection{\textbf{Service Caching for MEC Resource Allocation}}

Unlike the central cloud server that is always assumed with huge and diverse resources (e.g., computing, memory and storage),  the current edge server has much less resources, making it unable to accommodate all users' computation requests. On the other hand, different mobile services require different resources, based on which, they can be classified into CPU-hungry (e.g., cloud chess and VR), memory-hungry (e.g., online Matlab), and storage-hungry (e.g., VR) applications. Such a mismatch between resource and demand introduces a key challenge on how to allocate heterogeneous resources for service caching.

Note that similar problems have been investigated in conventional Cloud Computing systems\cite{tordsson2012cloud,li2009enacloud, gao2013multi,lucas2013scheduling}, termed as \emph{VM placement}, as well as MCC systems \cite{yang2016cost}. Specifically, the authors in \cite{tordsson2012cloud} proposed a novel architecture for VM management and optimized the VM placement over multiple clouds to reduce the deployment costs and improve user experience, given constraints on hardware configuration, the number of VMs as well as load balancing. Similar VM-placement problems were also investigated in \cite{li2009enacloud, gao2013multi} for maximizing the energy savings of cloud servers and in \cite{lucas2013scheduling} for different cloud scheduling strategies. Recently, the authors in \cite{yang2016cost} extended the VM placement idea to MCC systems and studied the joint optimization of service caching/placement over multiple clouds and load dispatching for end users' requests. As a result, one efficient algorithm was proposed to minimize both the computation latency and service placement transition cost. These works, however, cannot be directly applied to design efficient service caching policies for MEC systems, since it should take into account more refined information including users' location, preference, experience as well as edge servers' capacities in terms of the memory, storage and VM instance. To this end, two possible approaches are described as follows.

The first one is \emph{spatial popularity-driven service caching}, referring to caching different combinations and amounts of services in different MEC servers according to their specific locations and surrounding users' common interests. This idea is motivated by the fact that users in one small region are likely to request similar computing services. For example, visitors in a museum tend to use AR for better sensational experience. Thus, it is desirable to cache multiple AR services at the MEC server of this region for providing the real-time service. To achieve the optimal spatial service caching, it is essential to construct a \emph{spatial-application popularity distribution model} for characterizing the popularity of each application over different locations. Based on this, we can design resource-allocation policies using various optimization algorithms, e.g., the game theory and convex optimization techniques.

An alternative approach is \emph{temporal popularity-driven service caching}. The main idea is similar to that of the spatial counterpart, but it exploits the popularity information in the temporal domain, since the computation requests also depend on the time period. One example is that users are apt to play mobile cloud gaming after dinner. This kind of information will suggest MEC operators to cache several gaming services during this typical period for handling the huge computation loads. One disadvantage of this temporal-based approach is the additional server cost resulted from frequent \emph{cache-and-tear} operations since popularity information is time-varying and MEC servers possess finite resources.

\subsubsection{\textbf{Data Caching for MEC Data Analytics}}
Many modern mobile applications involve intensive computation based on data analytics, e.g., ranking and classification. Take VR as an instance. It creates an imaginary environment similar to the real world by generating realistic images, sounds and other sensations for enhancing users' experience. Achieving this end is nontrivial as it requires the MEC server to finish multiple complicated processes within the ultra-short duration (e.g., 1ms), such as recognizing users' actions via pattern recognition, ``understanding" users' requests via data mining, as well as rendering virtual settings via video streaming or other sensation techniques \cite{rheingold1991virtual}. All the above data-analytics based techniques should be supported by comprehensive database, which, however, imposes extremely heavy burden on the edge server storage. This challenge can be relieved by intelligent data caching that only reserves frequently-used database. From another perspective, caching parts of computation-result data that is likely to be reused by others can further boost the computation performance of the entire MEC system. One typical example is mobile cloud gaming, which enables fast and energy-efficient gaming by shifting game computing engines from mobiles to edge servers and supporting real-time gaming by game video streaming. Thus, it emerges as a leading technique for next generation mobile computing infrastructures \cite{wang2009modeling}. Since certain game rendered videos, e.g., gaming scenes, can be reused by other players, caching these computation results would not only significantly reduce the computation latency of the players with the same computation request, but also ease the computation burden for edge servers. Similar idea has been proposed in \cite{tran2016collaborative}, which investigated collaborative multi-bitrate video caching and processing in MEC.

For MEC data caching at a single edge server, one key problem is \emph{how to balance the tradeoff between massive database and finite storage capacity}. Unlike FemtoCaching networks where content (data) caching mainly introduces a new multiple-access mechanism termed as cache-enabled access \cite{bacstu2015cache}, data caching in MEC systems brings about manifold effects on the computation accuracy, latency and edge server-energy consumption, which, however, have not been characterized in existing literature. This calls for model building research efforts for accurately quantifying the mentioned effects for various MEC applications. Furthermore, it is also essential to establish a practical \emph{database popularity distribution model} that is able to statistically characterize the usage of each database set for different MEC applications. Based on the above models, the said tradeoff can be achieved by solving an optimization problem that maximizes the achievable QoS and minimizes the storage cost in MEC systems simultaneously.

The above framework can be further extended to MEC systems with multiple servers where each server can serve multiple users and each user can offload computation to multiple edge servers. The fundamental problem is similar to that of the cache-enabled HetNets \cite{cui2015analysis}, that is, how to \emph{spatially} distribute the database over heterogeneous edge servers under both storage and computation-load constraints on each of them, for increasing network-wide revenue. Intuitively, for each MEC server, it is desirable to spare more storage to cache the database of the most popular applications in its cell, and it also needs to utilize partial storage to accommodate less popular ones, whose computation performance will be further improved by cooperative caching in different MEC servers. Moreover, the performance of large-scale cache-enabled MEC networks can be analyzed using stochastic geometry by modeling nearby users as clusters \cite{suryaprakash2015modeling}.

\subsection{Mobility Management for MEC}
\begin{figure*}[!t]
\begin{center}
   \includegraphics[width=0.6\textwidth]{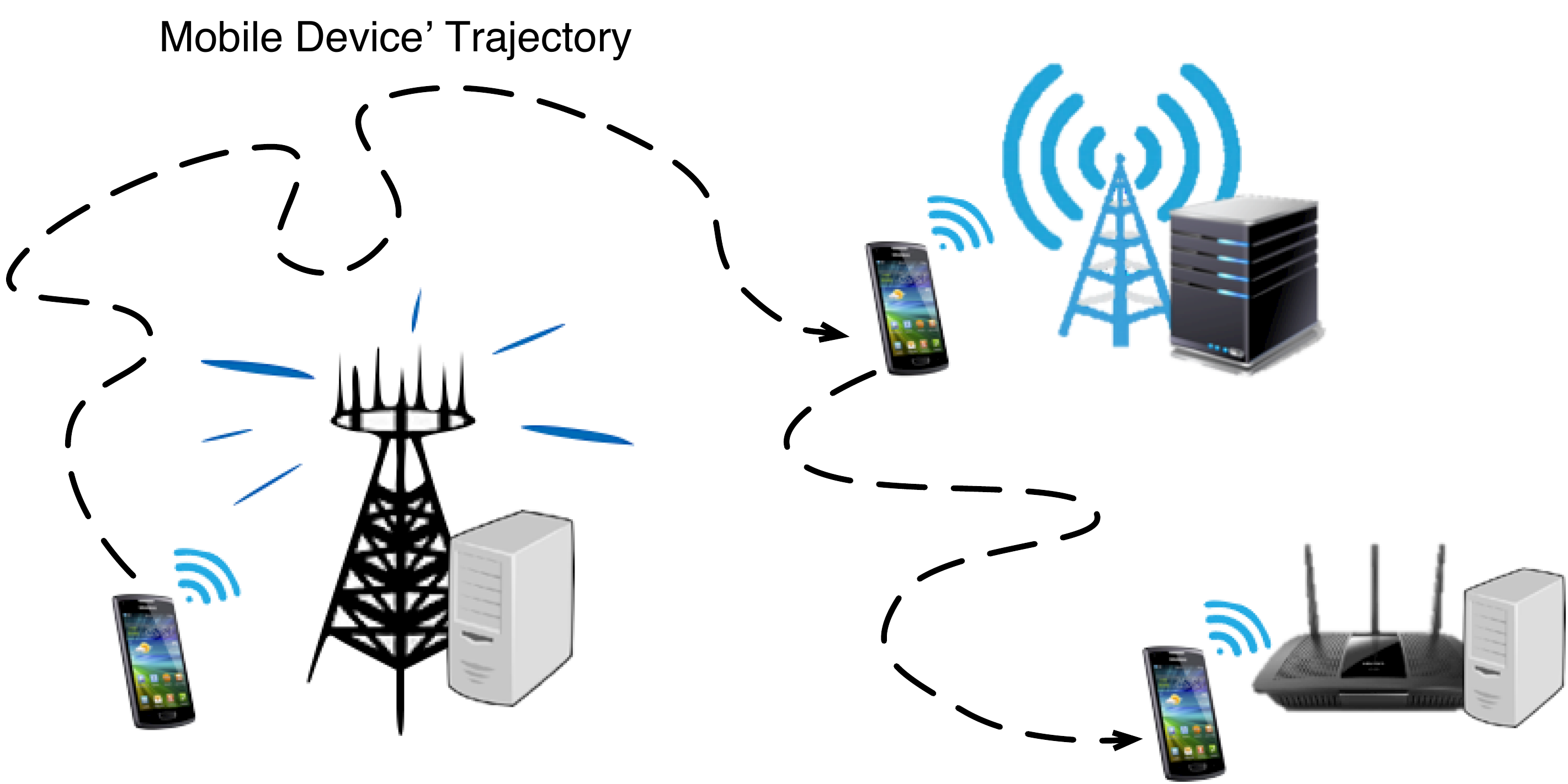}
\end{center}
\caption{Mobility management for MEC.}
\label{Mobility_management}
\end{figure*}

Mobility is an intrinsic trait of many MEC applications, such as VR assisted museum tour to enhance experience of visitors. In these applications, the movement and trajectory of users provide location and personal preference information for the edge servers to improve the efficiency of handling users' computation requests. On the other hand, mobility also poses significant challenges for realizing ubiquitous and reliable computing (i.e., without interruptions and errors) due to the following reasons. First, MEC will be typically implemented in the HetNet architecture comprising of multiple macro, small-cell BSs and WiFi APs. Thus, users' movement will call for frequent handovers among the small-coverage edge servers as shown in Fig.~\ref{Mobility_management}, which is highly complicated due to the diverse system configurations and user-server association policies. Next, users moving among different cells will incur severe interference and pilot contamination, which shall greatly degrade the communication performance. Last, frequent handovers will increase the computation latency and thus deteriorate users' experience.

Mobility management has been extensively studied for traditional heterogeneous cellular networks \cite{lopez2012mobility,damnjanovic2011survey,kassar2008overview}. In these prior works, users' mobility is modeled by the connectivity probability or the link reliability according to such information as the users' moving speeds. Based on such models, dynamic mobility management has been proposed to achieve high data rate and low bit-error rate. However, these policies cannot be directly applied for MEC systems with moving users, since they neglect the effects of the computation resources at edge servers on the handover policies. Recent works in \cite{wangCZ2014mobility,zhang2015offloading,lee2015user,rahimi2013music} have made initial efforts to design mobility-aware MEC systems. Specifically, the inter-contact time and contact rate were defined in \cite{wangCZ2014mobility} to model users' mobility. An opportunistic offloading policy was then designed by solving a convex optimization problem for maximizing the successful task offloading probability. Alternatively, to account for the mobility, the number of edge servers that users can access  was modeled by an HPPP in \cite{zhang2015offloading}. Then, the offloading decision was optimized by addressing the formulated MDP problem to minimize the offloading cost including mobile-energy consumption, latency and failure penalty. Other mobility models were also proposed in \cite{lee2015user,rahimi2013music}, which characterize the mobility by a sequence of networks that users can connect to and a two-dimensional location-time workflow, respectively. In addition, mobility management for MEC was integrated with traffic control in \cite{prasad2015efficient} to provide better experience for users with latency-tolerant tasks via designing intelligent cell association mechanisms. {\color{black}{In \cite{gomes2017edge}, edge caching was integrated with mobility prediction in Follow-Me Cloud for enhancing the content-caches migration located at the edges.}} Recent proposals on mobility-aware wireless caching in \cite{wang2016mobility} also provided valuable guidelines on mobility management in MEC systems.

Note that most of the existing works focused on optimizing mobility-aware server selection. However, to achieve better user experience and higher network-wide profit, the offloading techniques at mobile devices and scheduling policies at MEC servers should be jointly considered. This introduces a set of interesting research opportunities with some described as follows.

\subsubsection{\textbf{Mobility-Aware Online Prefetching}} In practice, the full information of the user trajectory may be unavailable. Conventional design for mobile computation offloading will fetch a computation task to another server only when it is handoverred. This mechanism requires excessive fetching of a large volume of data for handover and thus brings long fetching latency. Moreover, it also causes heavy loads on the MEC network. One promising solution to handle this issue is to leverage the statistical information of the user trajectory and prefetch parts of future computation data to potential servers during the server-computation time, referred to as \emph{online prefetching} \cite{ko2016online}. This technique can not only significantly reduce the handover latency via mobility prediction, but also enable energy-efficient computation offloading by enlarging the transmission time. However, it also encounters several challenges with two most critical ones described as follows. The first challenge arises from the trajectory prediction. Accurate prediction can allow seamless handovers among edge servers and reduce the prefetching redundancy. Achieving it, however, requires precise modeling and high-complexity ML techniques, e.g., Bayesian, reinforcement and deep learning. For example, the trajectory of a typical visitor in a museum can be predicted according to his own interest-information and statistical route-information of some previous visitors with similar interests that can be obtained by ML algorithms. Therefore, it is important to balance the tradeoff between the modeling accuracy and computation complexity. The second challenge lies in the selection of the prefetched computation data. To maximize the successful offloading probability of edge users, the computation-intensive components should be prefetched earlier with adaptive transmission power control in dynamic fading channels.

\subsubsection{\textbf{Mobility-Aware Offloading Using D2D Communications}}
D2D communications was first proposed in \cite{doppler2009device} to improve the network capacity and alleviate the data traffic burden in cellular systems. This paradigm can also be used to handle the user mobility problems in MEC systems \cite{li2014exploring}, which creates numerous D2D communication links. These links allow the computation of a user to be offloaded to its nearby users which have more powerful computation capabilities. The short-range communication offered by D2D links reduces energy consumption of data transmission as well. However, user mobility brings new  design issues as follows. The first one is how to exploit the advantages of both D2D and cellular communications. One possible approach is to offload the computation-intensive data to the edge servers at BSs that have huge computation capabilities in order to reduce the server-computing time; while the components of large data sizes and strict computation requirements should be fetched to nearby users via D2D communications for higher energy efficiency. Next, the selection of surrounding users for offloading should be optimized to account for users' mobility information, dynamic channels and heterogeneous users' computation capabilities. Last, massive D2D links will introduce severe interference for reliable communications. This issue is more complicated in the mobility-based MEC systems due to the fast-changing wireless fading environments. Hence, advanced interference cancellation and cognitive radio techniques can be applied for MEC systems, together with mobility prediction to increase the offloading rate and reduce the service latency.

\subsubsection{\textbf{Mobility-Aware Fault-Tolerant MEC}} User mobility poses significant challenges for providing reliable MEC services due to dynamic environments. Computation offloading may fail due to intermittent connections and rapid-changing wireless channels. The induced failure is catastrophic for the latency-sensitive and resource-demanding applications. For instance, AR-based museum video guide aims to provide fluent and fancy virtual sensations for visitors, and the disruption or failure of video streaming due to intermittent connections would upset visitors. Another example is the military operation which always requires fast and ultra-reliable computation, even in high-mobility environments. Any computation failure would bring serious consequences. These facts necessitate the design for mobility-aware fault-tolerant MEC systems \cite{chen2015energy,chenenergy,satria2017recovery}, with three major and interesting problems illustrated as follows, including fault prevention, fault detection and fault recovery. Fault prevention is to avoid or prevent MEC fault by backing up extra stable offloading links. Macro BSs or central clouds can  be chosen as protection-clouds, since they have large network coverage that allows continuous MEC service. The key design challenges lie in how to balance the tradeoff between QoS (i.e., the failure probability) and energy consumption due to extra offloading links for the single-user case, and how to allocate protection-clouds for multiuser MEC applications. Next, fault detection is to collect fault information, which can be realized by setting intelligent timing checks or receiving feedbacks for MEC services. In addition, channel and mobility estimation techniques can also be applied to estimate the fault so as to reduce the detection time. Last, for detected MEC faults, recovery approaches should be performed to continue and accelerate the MEC service. The suspended service can be switched to more reliable backup wireless links with adaptive power control for higher-speed offloading. {\color{black}{Alternative recovery approaches include migrating the workloads to neighboring MEC systems directly or through ad-hoc relay nodes as proposed in \cite{satria2017recovery}. }}

\subsubsection{\textbf{Mobility-Aware Server Scheduling}} For multiuser MEC systems, traditional MEC server scheduling servers users according to the offloading priority order that depends on users' distinct local computing information, channel gains and latency requirements \cite{you2016energy}. However, this static scheduling design cannot be directly applied for the multiuser MEC systems with mobility due to dynamic environments, e.g., time-varying channels and intermittent connectivities. Such dynamics motivate the design of adaptive server scheduling that regenerates the scheduling order from time to time, incorporating the real-time user information. In such adaptive scheduling mechanisms, users with worse conditions will be allocated with higher offloading priorities to meet their computing deadlines. Another potential approach is to design mobility-aware offloading priority function by the following two steps. The first step is to accurately predict users' mobility profiles and channels, where the major challenge is how to reflect the mobility effects and re-define the offloading priority function. The second step is resource reservation that can enhance the server scheduling performance \cite{chaisiri2012optimization,zhang2016reservation}. Specifically, to guarantee the QoS of latency-sensitive and high-mobility users,  MEC servers can reserve some dedicated computational resources and provide reliable computing service for such users. While for other latency-tolerant users, the MEC server can perform on-demand provisioning. For such a hybrid MEC server provisioning scheme, the server scheduling can be optimized for serving the maximum number of users with QoS guarantees, as well as maximizing MEC servers' revenue.

\subsection{Green MEC}

MEC servers are small-scale data centers, each of which consumes substantially less energy than the conventional cloud data center. However, their dense deployment pattern raises a big concern on the system-wide energy consumption. Therefore, it is unquestionably important to develop innovative techniques for achieving green MEC \cite{XJinGreenDCSurvey16,XSunIEEENetw1602}. Unfortunately, designing green MEC is much more challenging compared to green communication systems or green DCNs. Compared to green communication systems, the computational resource needs to be managed to guarantee satisfactory computation performance, making the traditional green radio techniques not readily applicable. On the other hand, the previous research efforts on green DCNs have not considered the radio resource management, which makes them not suitable for green MEC. Besides, the highly unpredictable computation workload pattern in MEC servers poses another big challenge for resource management in MEC systems, calling for advanced estimation and optimization techniques. In this subsection, we will introduce different approaches on designing green MEC systems, including dynamic right-sizing for energy-proportional MEC, \emph{geographical load balancing} (GLB) for MEC, and MEC systems powered by renewable energy.

\subsubsection{\textbf{Dynamic Right-Sizing for Energy-Proportional MEC}}

The energy consumption of an MEC server highly depends the utilization radio [see Eq. (\ref{ServerEnergyUtiRatio})]. Even when the server is idling, it still consumes around 70\% of the energy as it operates at the full speed. This fact motivates the  design of \emph{energy-proportional} (or \emph{power-proportional}) servers, i.e., the energy consumption of a server should be proportional to its computation load \cite{Barroso07}. One way to realize energy-proportional servers is to switch off/slow down the processing speeds of some edge servers with light computation loads. Such an operation is termed as \emph{dynamic right-sizing} in the literature on green DCNs \cite{MLin1310}. However, along with the potential energy savings, toggling servers between the active and sleep modes could bring detrimental effects. First of all, it will incur the switching energy cost and application data-migration latency. Also, user experience may be degraded due to the less amount of allocated computational resources, which may, in turn, reduce the operator's revenue. Besides, the risk associated with server toggling as well as the \emph{wear-and-tear} cost of the servers might be increased, which can in turn increase the maintenance costs of MEC vendors. As a result, switching off the edge servers in a myopic manner is not always beneficial.

In order to make an effective decision on dynamic right-sizing, the profile of computation workload at each edge server should be accurately forecasted. In conventional DCNs, this can be achieved rather easily as the workload at each data center is an aggregation of the computation requests across a large physical region, e.g., several states in the United States, which is relatively stable so that it can be estimated by referring to the readily available historical data at the data centers. However, for MEC systems, the serving area of each edge server is much smaller, and hence its workload pattern is affected by many factors, such as the location of the server, time, weather, the number of nearby edge servers, and user mobility. This leads to a fast-changing workload pattern, and requires more advanced prediction techniques. Moreover, online dynamic right-sizing algorithms that require less future information need to be developed.

\subsubsection{\textbf{Geographical Load Balancing for MEC}}

GLB is another key technique for green DCNs \cite{MLin1206,HXu1506TPDS}, which leverages the \emph{spatial diversities} of the workload patterns, temperatures, and electricity prices, to make workload routing decision among different data centers. This technique can also be applied to MEC systems. For instance, a cluster of MEC servers can coordinate together to serve a mobile user, i.e., the tasks can be routed from the edge server located in a hot spot (such as a restaurant) to a nearby edge server with light workload (such as the one in a park). On one hand, this helps to improve the energy efficiency of the lightly-loaded edge servers as well as user experience. On the other hand, it can prolong the battery lives of mobile devices, as offloading the tasks through the nearby server could save transmission energy. It is worthwhile to note that the implementation of GLB requires efficient resource management techniques at edge servers, such as dynamic right-sizing and VM management \cite{BeloglazovCCGrid1005,XLiINFOCOM1404,LChenINFOCOM1404,ZHanINFOCOM1604}.

Meanwhile, there are many factors to be incorporated when applying GLB in MEC environments. Firstly, since the migrated tasks should go through the cellular core network, the network congestion state should be monitored and considered when making the GLB decisions. Secondly, to enable seamless task migration, a VM should be migrated/set up in another edge server beforehand, which may cause additional energy consumption. Thirdly, the mutual interests of MEC operators and edge computing service subscribers should be carefully considered when performing GLB, due to the tradeoff between the energy savings and latency reduction. Last but not least, the existence of conventional Cloud Computing infrastructures endows the edge servers with an extra option of offloading the latency-critical and computation-intensive tasks to remote cloud data centers, creating a new design dimension and further complicating the optimization.

\subsubsection{\textbf{Renewable Energy-Powered MEC Systems}}
\begin{figure*}[!t]
\begin{center}
   \includegraphics[width=0.6\textwidth]{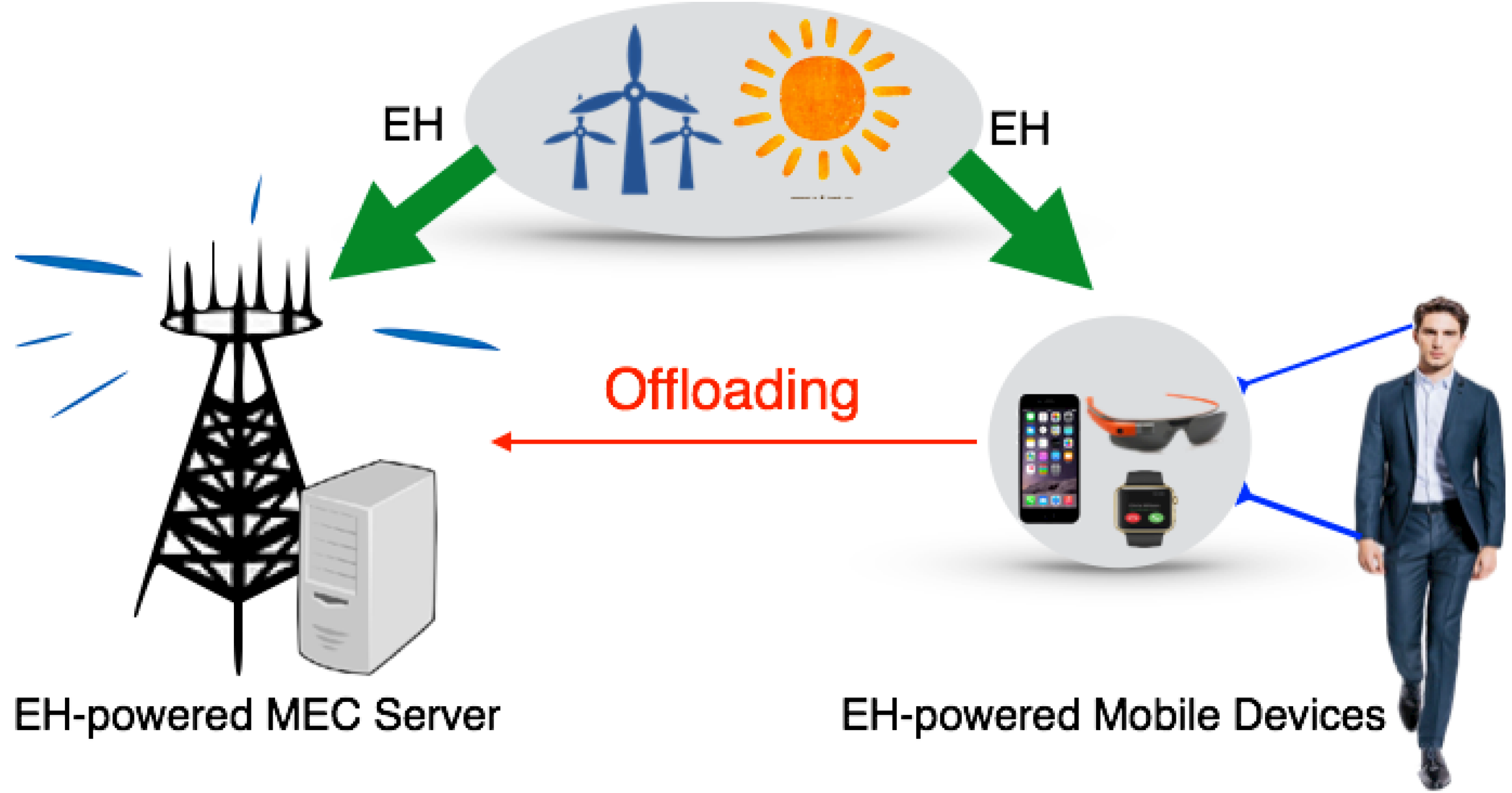}
\end{center}
\caption{Renewable energy-powered MEC systems.}
\label{EH_MEC}
\end{figure*}
Traditional grid energy is normally generated by coal-fired power plants. Hence, powering mobile systems with grid energy inevitably causes a huge amount of carbon emission, which opposes the target of green computing. Off-grid renewable energy, such as solar radiation and wind energy, recently, has emerged as a viable and promising power source for various IT systems thanks to the recent advancements of \emph{energy harvesting} (EH) techniques \cite{SudevalayamSandT11,SUlukusJSAC1503}. This fact motivates the design of innovative MEC systems, called renewable energy-powered MEC systems, which are shown in Fig. \ref{EH_MEC} comprising both EH-powered MEC servers and mobile devices.
On one hand, as the MEC servers are expected to be densely-deployed and have low power consumption similar to that of small-cell BSs \cite{YMaoComMag1506}, it is reasonable and feasible to power the MEC infrastructures with the state-of-the-art EH techniques. On the other hand, the mobile devices can also get benefits from using renewable energy as EH is able to prolong their battery lives, which is one of the most favorable features for mobile phones \cite{CNNbattery05}. Besides, the use of renewable energy sources eliminates the need of human intervention such as replacing/recharging the batteries, which is difficult if not impossible for certain types of application scenarios where the devices are hard and dangerous to reach. Meanwhile, these advantages of using renewable energy are accompanied with new design challenges.

A fundamental problem to be addressed for renewable energy-powered MEC systems is the \emph{green energy-aware resource allocation and computation offloading}. Instead of minimizing the energy consumption subject to satisfactory user experience, the design principle for the renewable energy-powered MEC systems should be changed to optimizing the achievable performance given the renewable energy constraint, as the renewable energy almost comes for free. Also, with renewable energy supplies, the \emph{energy side information} (ESI), which indicates the amount of available renewable energy, will play a key role in the decision making. Initial investigations on renewable energy-powered MEC systems were conducted in \cite{JXu1612GC} and \cite{YMao1612JSAC}, which focused on EH-powered MEC servers and EH-powered mobile devices, respectively. For EH-powered MEC servers, the system operator should decide the amount of workload required to be offloaded from the edge server to the central cloud, as well as the processing speed of the edge server, according to the information of the core network congestion state, computation workload, and ESI. This problem was solved by a learning-based online algorithm in \cite{JXu1612GC}. While for EH-powered mobile devices, a dynamic computation offloading policy has been proposed in \cite{YMao1612JSAC} using Lyapunov optimization techniques based on both the CSI and ESI. However, these two works only considered small-scale MEC systems that consist of either one edge server (in \cite{JXu1612GC}) or one mobile device (in \cite{YMao1612JSAC}). Thus, they cannot provide a comprehensive solution for large-scale MEC systems.

For large-scale MEC systems where multiple MEC servers are deployed across a large geographic region, the concept of GLB could be modified as the \emph{green energy-aware GLB} to optimize the MEC systems by further utilizing the spatial diversity of the available renewable energy. This idea was originally proposed for green DCNs, where the \emph{``follow the renewables''} routing scheme offers a huge opportunity in reducing the grid energy consumption \cite{MLin1206,CChenCloudCom1212,CDongIGCC1306,XSunCloudCom1511,TChenJSAC1603}. Moreover, as mentioned before, there exist significant differences between MEC systems and conventional DCNs in terms of the wireless channel fluctuation and resource-management design freedom of system operators. These factors make the offloading decision making for the green energy-aware GLB in MEC systems much more complicated, as it needs to consider the CSI and ESI in the whole system.

The randomness of renewable energy may introduce the offloading unreliability and risks of failure, bringing about a major concern for using renewable energy to power MEC systems. Fortunately, there are several potential solutions to circumvent this issue as described below.

\begin{itemize}
\item First, thanks to the low deployment cost, renewable energy-powered edge servers can be densely deployed over the system to provide more offloading opportunities for the users.  The resultant overlapping serving areas offer the offloading diversity in the available energy to avoid performance degradation. A similar idea has been proposed for EH cooperative communication systems in \cite{YLuoTCOM1603}.

\item {Second, the chance of energy shortage can be reduced by properly selecting the renewable energy sources. It was found in \cite{MLin1206} that solar energy is more suitable for workloads with a high \emph{peak-to-mean ratio} (PMR), while wind energy fits better for workloads with a small PMR. This provides guidelines for renewable energy provisioning for edge servers.}

\item Third, MEC servers can be powered by hybrid energy sources to improve reliability \cite{JGongTCOM13,THanTWC1309,YMaoTWC16}, i.e., powered by both the electric grid and the harvested energy. Also, equipping \emph{uninterrupted power supply} (UPS) units at the edge servers can provide a short period of stable energy supply when green energy is in deficit, and it can be recharged when the surrounding energy condition returns to a good state.

\item Moreover, \emph{wireless power transfer} (WPT), which charges mobile devices using RF wave \cite{Brown198409,HJuTWC1401}, is a newly-emerged solution that enables wireless charging and extends the battery life. This technique has been provided in modern mobile phones such as Samsung Galaxy S6. In renewable energy-powered MEC systems, the edge servers can be powered by WPT when the renewable energy is insufficient for reliability \cite{HKBTWC14}. This technology also applies to the computation offloading for mobile devices in MEC systems \cite{you2016energyJSAC} and data offloading for collaborate mobile clouds \cite{chang2016energy}. However, novel energy beamforming techniques are needed to increase the charging efficiency. Moreover, due to the double near-far problem in wireless powered systems, it requires a delicate scheduling to guarantee fairness among multiple mobile devices.
\end{itemize}

\subsection{Security and Privacy Issues in MEC}

There are increasing demands for secure and privacy-preserving mobile services. While MEC enables new types of services, its unique features also bring new security and privacy issues. First of all, the innate heterogeneity of MEC systems makes the conventional trust and authentication mechanisms inapplicable. Second, the diversity of communication technologies that support MEC and the software nature of the networking management mechanisms bring new security threats. Besides, secure and private computation mechanisms become highly desirable as the edge servers may be an eavesdropper or an attacker. These motivate us to develop effective mechanisms as described in the following.

\subsubsection{\textbf{Trust and Authentication Mechanisms}}

Trust is an important security mechanism in almost every mobile system, behind which, the basic idea is \emph{to know the identity of the entity that the system is interacting with}. Authentication management provides a possible solution to ensure ``trust'' \cite{RomanLM16}. However, the inherent heterogeneity of MEC systems, i.e., different types of edge servers may be deployed by multiple vendors and different kinds of mobile devices coexist, makes the conventional trust and authentication mechanisms designed for Cloud Computing systems inapplicable. For example, the reputation-based trust model will lead to severe trust threats in MEC systems, as demonstrated in \cite{SYiWASA1508}. This fact calls for a unified trust and authentication mechanism that is able to assess the reliability of edge servers and identify the camouflaged edge servers. Besides, within the mobile network, there will be a large number of edge servers serving massive mobile devices. This makes the trust and authentication mechanism design much more complicated compared with that in conventional Cloud Computing systems, since edge servers are of small computation capabilities and designed to enable latency-sensitive applications. Therefore, it is critical to minimize the overhead of authentication mechanisms and design distributed policies \cite{FoudaTSG1112,AhmedIACC1302}.

\subsubsection{\textbf{Networking Security}}

The communication technologies to support MEC systems, e.g., WiFi, LTE and 5G, have their own security protocols to protect the system from attacks and intrusions. However, these protocols inevitably create different trust domains. The first challenge of networking security in MEC systems comes from the difficulties in the distribution of credentials, which can be used to negotiate session keys among different trust domains \cite{RomanLM16}. In existing solutions, the certification authority can only distribute the credentials to all the elements located within its own trust domain \cite{RomanLM16}, making it hard to guarantee the privacy and data integrity for communications among different trust domains. To address this problem, we can use the cryptographic attributes as credentials in order to exchange session keys \cite{XHuang14TDSC,Gorantla1007}. Also, the concept of federated content networks, which defines how multiple trust domains can negotiate and maintain inter-domain credentials \cite{PimentelComptCommun1509}, can be utilized.

Besides, techniques such as SDN and NFV are introduced to MEC systems to simplify the networking management as well as to provide isolation \cite{ETSI14}. However, these techniques are softwares by nature and thus vulnerable \cite{Liyanage1607,WYang1606}. Moreover, the large number of devices and entities in MEC systems increase the chance of successfully attacking a single device, which provides means to launch an attack to the whole system \cite{BLiang17}. Therefore, novel and robust security mechanisms, such as hypervisor introspection, run-time memory analysis, and centralized security management \cite{AlcatelNFVsecurity}, are needed to guarantee a secured networking environment for MEC systems.

\subsubsection{\textbf{Secure and Private Computation}}

Migrating computation-intensive applications to the edge servers is the most important function and motivation of building MEC systems. In practice, the task input data commonly contains sensitive and private information such as personal clinical data and business financial records. Therefore, such data should be properly pre-processed before being offloaded to edge servers, especially the untrusted ones, in order to avoid information leakage. In addition to information leakage, the edge servers may return inaccurate and even incorrect computation results  due to either software bugs or financial incentives, especially for tasks with huge computation demands \cite{CWang1601}. To achieve secure and private computation, it is highly preferred that the edge platforms can execute the computation tasks without the need of knowing the original user data and the correctness of the computation results can be verified, which can be realized by encryption algorithms and verifiable computing techniques \cite{Gennaro10}. An interesting example of secure computation mechanisms for LP problems was developed in \cite{CWang1601}, where the LP problem is decomposed into the public-owned solvers and the private-owned data. By using a privacy-preserving transformation, the customer offloads the encrypted private data for cloud execution, and the server returns the results for the transformed LP problem. A set of necessary and sufficient conditions for verifying the correctness of the results were developed based on duality theory. Upon receiving the correct result, the clients can map back the desired solution for the original problem using the secret transformation. This method of result validation achieves a big improvement in computation efficiency via high-level LP computation compared to the generic circuit representation, and it incurs close-to-zero additional overhead on both the client and cloud server, which provides hints to develop secure and private computation mechanisms for other cloud applications.
\begin{figure*}[!t]
\begin{center}
   \includegraphics[width=0.7\textwidth]{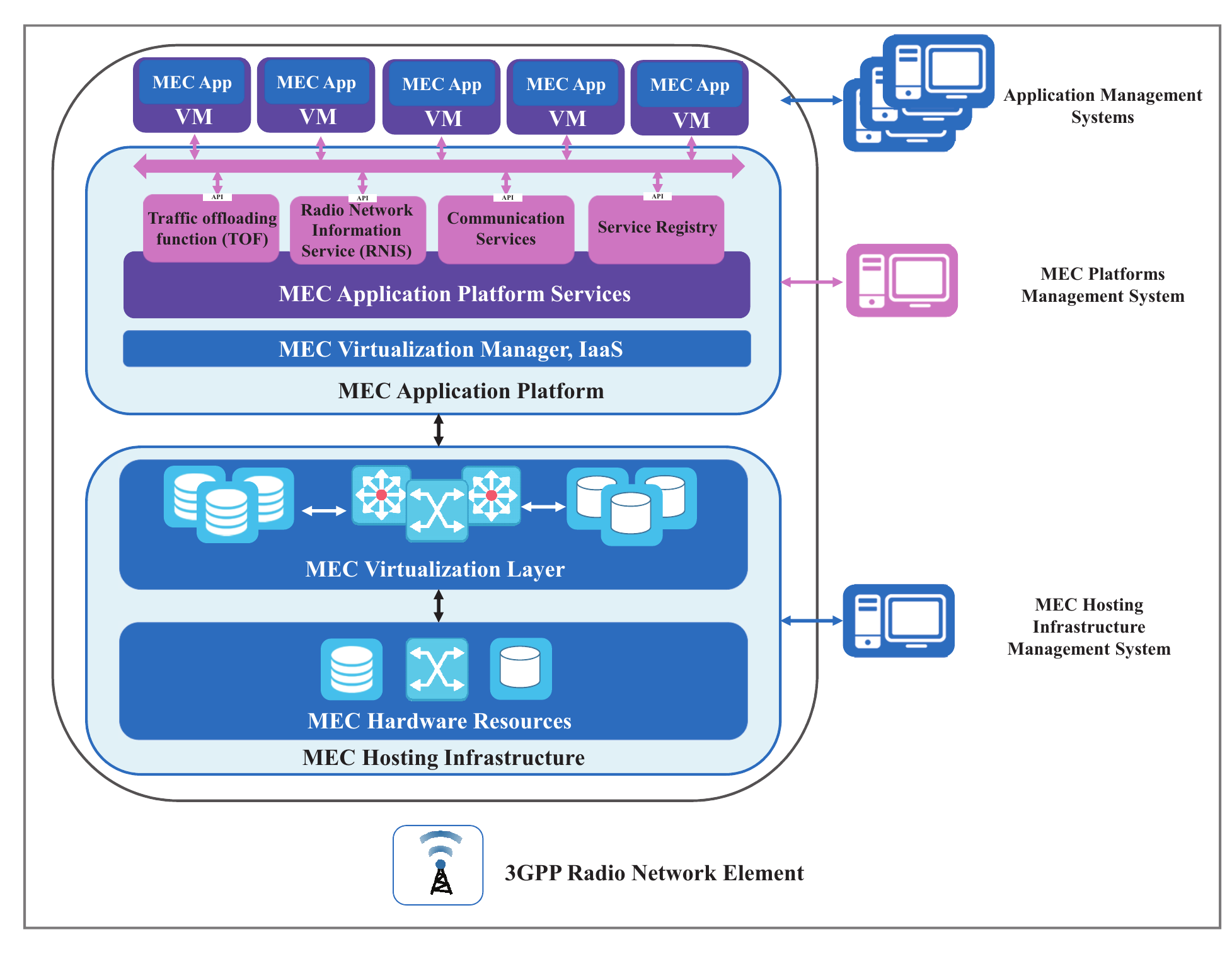}
\end{center}
\caption{MEC platform overview \cite{ETSI14}.}
\label{MECplatformOverview}
\end{figure*}
\section{Standardization Efforts and Use Scenarios of MEC}\label{Section:Standard}

Standardization is an indispensable step for successful promotion of a new technology, which documents the consensus among multiple players and defines voluntary characteristics and rules in a specific industry. Due to the availability of structured methods and reliable data, standardization helps to promote innovation and disseminate groupbreaking ideas and knowledge about cutting-edge techniques. More importantly, standardization can build customer trust in products, services and systems, which helps to develop favorable market condition. The technical standards for MEC are being developed by ETSI, and a new \emph{industry specification group} (ISG) was established within ETSI by Huawei, IBM, Nokia Networks, NTT docomo and Vodafone. The aim of the ISG is to build up a standardized and open environment, which will allow the efficient and seamless integration of applications from vendors, service providers, and third-parties across multi-vendor MEC platforms \cite{ETSI_Briefing}. In September 2014, an introductory technical white paper on MEC was published by ETSI, which defined the concept of MEC, proposed the referenced MEC platform, as well as pointed out a set of technical requirements and challenges for MEC \cite{ETSI14}. Also, typical use scenarios and their relationships with MEC have been discussed. These aspects have also been documented in the ETSI specifications in 2015 \cite{ETSI_Term,ETSI_ServiceScenarios,ETSI_Framework,ETSI_TechRequir}. Most recently, ETSI has announced six \emph{Proofs of Concepts} (PoCs) that were accepted by the MEC ISG in MEC World Congress 2016, which will assist the strategic planning and decision-making of organizations, as well as help to identify which MEC solutions may be viable in the network \cite{PoC1stETSI}. This provides the community with confidence in MEC and will accelerate the pace of the standardization. It is interesting to note that, in this congress, the ETSI MEC ISG has renamed \emph{Mobile Edge Computing} as \emph{Multi-access Edge Computing} in order to reflect the growing interest in MEC from non-cellular operators, which will take effects starting from 2017 \cite{MECmultipleaccess}. {\color{black}Most recently, the \emph{3rd Generation Partnership Project} (3GPP) shows a growing interest in including MEC into its 5G standard, and functionality supports for edge computing has been identified and reported in a recent technical specification document \cite{3GPPTS23501}.} In this section, we will first introduce the recent standardization efforts from the industry, including the referenced MEC server framework as well as the technical challenges and requirements of MEC systems. Typical use scenarios of MEC will be then elaborated. {\color{black}In addition, we will discuss MEC-related issues in 5G standardizations, including the functionality supports for MEC, and the innovative features in 5G systems with the potential to help realize MEC.}

\subsection{Referenced MEC Server Framework}

In the MEC introductory technical white paper \cite{ETSI14}, the ETSI MEC ISG has defined a referenced framework for MEC servers (a.k.a. MEC platforms), where each server consists of a hosting infrastructure and an application platform as shown in Fig. \ref{MECplatformOverview}. The hosting infrastructure includes the hardware components (such as the computation, memory, and networking resources) and an MEC virtualization layer (which abstracts the detailed hardware implementation to the MEC application platform). Also, the MEC host infrastructure provides the interface to the host infrastructure management system as well as the radio network elements, which, however, are beyond the scope of the MEC initiative due to the availability of multiple implementation options.

The MEC application platform includes an MEC virtualization manager together with an \emph{Infrastructure as a Service (IaaS)} controller, and provides multiple MEC application platform services. The MEC virtualization manager supports a hosting environment by providing IaaS facilities, while the IaaS controller provides a security and resource sandbox (i.e., a virtual environment) for both the applications and MEC platform. The MEC application platform offers four main categories of services, i.e., \emph{traffic offloading function} (TOF), \emph{radio network information services} (RNIS), communication services, and service registry. An MEC application platform management interface is used by the operators for MEC application platform management, supporting the application configuration and life cycle control, as well as VM operation management.

On top of the MEC application platform, the MEC applications are deployed and executed within the VMs, which are managed by their related application management systems and agnostic to the MEC server/platform and other MEC applications.

\subsection{Technical Challenges and Requirements}

In this subsection, we will briefly summarize the technical challenges and requirements specified in \cite{ETSI14,ETSI_TechRequir}.

\subsubsection{\textbf{Network Integration}} As MEC is a new type of service deployed on top of the communication networks, the MEC platform is supposed to be transparent to the 3GPP network architectures, i.e., the existing 3GPP specifications should not be largely affected by the introduction of MEC.

\subsubsection{\textbf{Application Portability}} Application portability requires MEC applications to be seamlessly loaded and executed by the MEC servers deployed by multiple vendors. This eliminates the need for dedicated development or integration efforts for each MEC platform, and provides more freedom on optimizing the location and execution of MEC applications. It requires the consistency of the MEC application platform management systems, as well as mechanisms used to package, deploy and manage applications from different platforms and vendors.

\subsubsection{\textbf{Security}} The MEC systems face more security challenges than communication networks due to the integration of computing and IT services. Hence, the security requirements for the 3GPP networks and the IT applications (e.g., isolating different applications as much as possible) should be simultaneously satisfied. Besides, because of the nature of proximity, the physical security of the MEC servers is more vulnerable compared to conventional data centers. Thus, the MEC platforms need to be designed in a way that both logical intrusions and physical intrusions are well protected. Moreover, authorization is an important aspect to prevent the unauthorized/untrusted third-party applications from destroying MEC hosts as well as the valued radio access network.

\subsubsection{\textbf{Performance}} As mentioned previously, the telecom operators expect that introducing MEC will have minimal impacts on the network performance, e.g., the throughput, latency, and packet loss. Thus, sufficient capacity should be provisioned to process the user traffic in the system deployment stage. Also, because of the highly-virtualized nature, the provided performance may be impaired especially for those applications that require intensive use of hardware resources or have low latency requirements. As a result, how to improve the efficiency of virtualized environments becomes a big challenge.

\subsubsection{\textbf{Resilience}} The MEC systems should offer certain level of resilience and meet the high-availability requirements demanded by their network operators. The MEC platforms and applications should have  fault-tolerant abilities to prevent them from adversely affecting other normal operations of the network.

\subsubsection{\textbf{Operation}} The virtualization and Cloud technologies make it possible for various parties to participate in the management of MEC systems. Thus, the implementation of the management framework should also consider the diversity of potential deployments.

\subsubsection{\textbf{Regulatory and Legal Considerations}} The development of MEC systems should meet the regulatory and legal requirements, e.g., the privacy and charging.

Besides the aforementioned challenges and requirements, there still exist more aspects that should be considered in the final MEC standards, such as the support for user mobility, applications/traffic migration, and requirements on the connectivity and storage. However, currently, the standardization efforts and even efforts from the research communities are still on their infant stages.
\subsection{Use Scenarios}
\label{UseScenarios}
MEC will enable numerous mobile applications. In this subsection, we will introduce four typical use scenarios that have been documented by ETSI MEC ISG in \cite{ETSI_ServiceScenarios}.

\subsubsection{\textbf{Video Stream Analysis Service}}
\begin{figure*}[!h]
\begin{center}
   \includegraphics[width=0.65\textwidth]{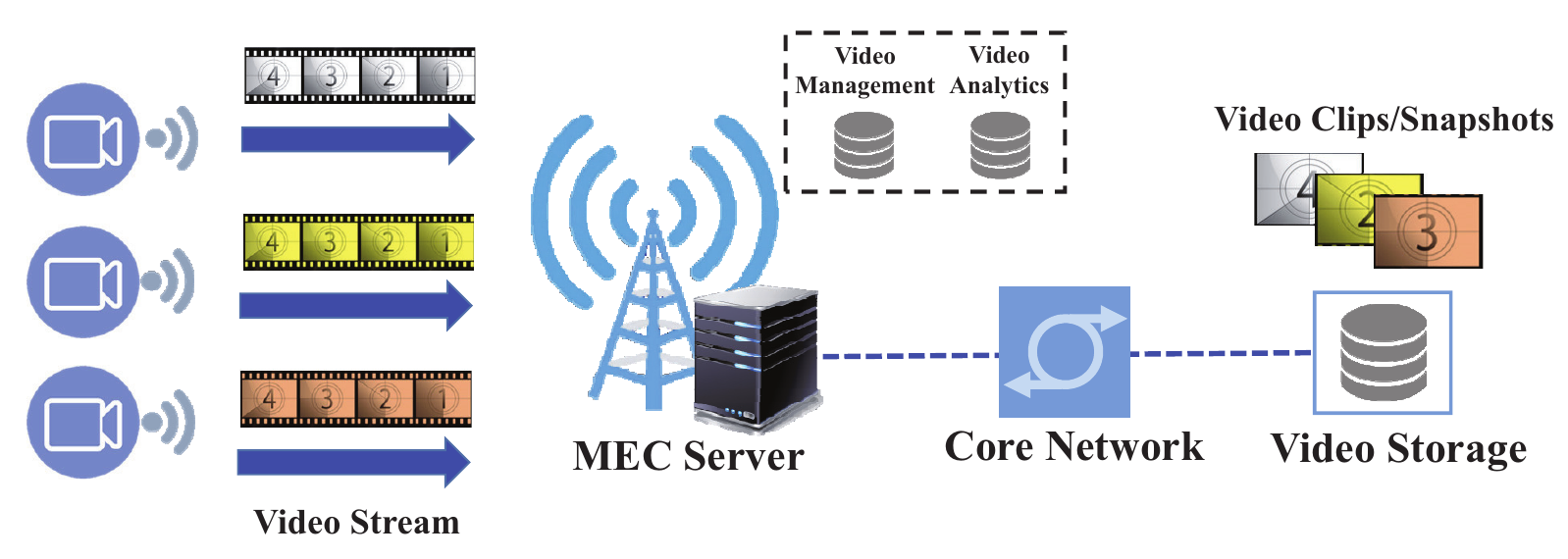}
\end{center}
\caption{MEC for video stream analysis \cite{ETSI14}.}
\label{UseScnVideo}
\end{figure*}
\begin{figure*}[!t]
\begin{center}
   \includegraphics[width=0.7\textwidth]{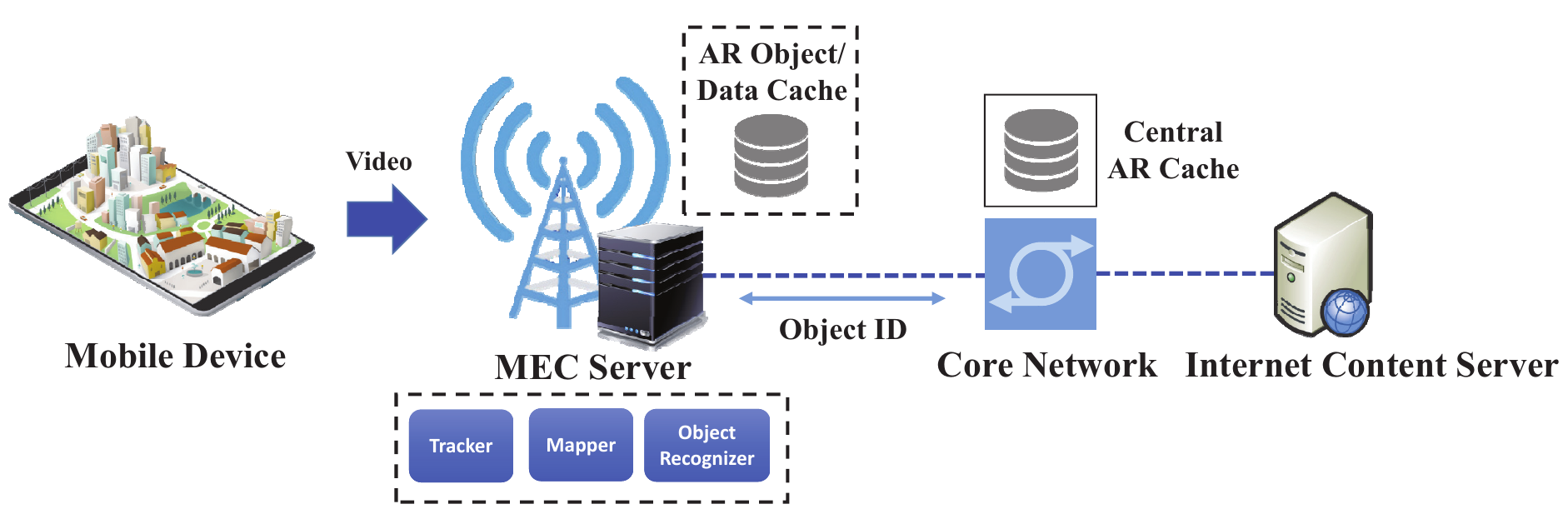}
\end{center}
\caption{MEC for AR services \cite{ETSI14}.}
\label{UseScnAR}
\end{figure*}

Video stream analysis has a broad range of applications such as the vehicular license plate recognition, face recognition, and home security surveillance, for which, the basic operations include object detection and classification. The video analysis algorithms normally have a high computation complexity, and thus it is preferable to move the analysis jobs away from the video-capturing devices (e.g., the camera) to simplify the device design and reduce the cost. If these processing tasks are handled in the central cloud, the video stream should be routed to the core network \cite{AnjumTCC16}, which will consume a great amount of network bandwidth due to the nature of video stream. By performing the video analysis in the place close to edge devices, the system can not only enjoy the benefits of low latency, but also avoid the problem of network congestion caused by the video stream uploading. The MEC-based video analysis system is shown in Fig. \ref{UseScnVideo}, where the edge server should have the ability to conduct video management and analysis, and only the valuable video clips (screenshots) will be backed up to the cloud data centers.

\subsubsection{\textbf{Augmented Reality Service}}

AR is a live direct or indirect view of a physical, real-world environment whose elements are augmented (or supplemented) by computer-generated sensory inputs such as sound, video, graphics, or GPS data\footnote{\url{https://en.wikipedia.org/wiki/Augmented_reality}}. Upon analyzing such information, the AR applications can provide additional information in real-time. The AR applications are highly localized and require low latency as well as intensive data processing. One of the most popular applications is the museum video guides, i.e., a handheld mobile device that provides the detailed information of some exhibits that cannot be easily shown on the scene. Online games, such as the Pok\'emon Go\footnote{http://www.pokemongo.com/}, is another important application that AR techniques play a critical role. An MEC-based AR application system is shown in Fig. \ref{UseScnAR}, where the MEC server should be able to distinguish the requested contents by accurately analyzing the input data, and then transmit the AR data back to the end user. Much attention has been paid on the MEC-enabled AR systems recently, and one demo has been implemented by Intel and roadshowed in the Mobile World Congress 2016 \cite{IntelMECCongressAR}.

\subsubsection{\textbf{IoT Applications}}
To simplify the hardware complexity of IoT devices and prolong their battery lives, it is promising to offload the computation-intensive tasks for remote processing and retrieve the results (required action) once the processing is completed. Also, some IoT applications need to obtain distributed information for computation, which might be difficult for an IoT device without the aid of an external entity. Since the MEC servers host high-performance computation capabilities and are able to collect distributed information, their deployment will significantly simplify the design of IoT devices, without the need to have strong processing power and capability to receive information from multiple sources for performing meaningful computation. Another important feature of IoT is the heterogeneity of the devices running different forms of protocols, and their management should be accomplished by a low-latency aggregation point (gateway), which could be the MEC server.

\subsubsection{\textbf{Connected Vehicles}}
\begin{figure*}[!t]
\begin{center}
   \includegraphics[width=0.75\textwidth]{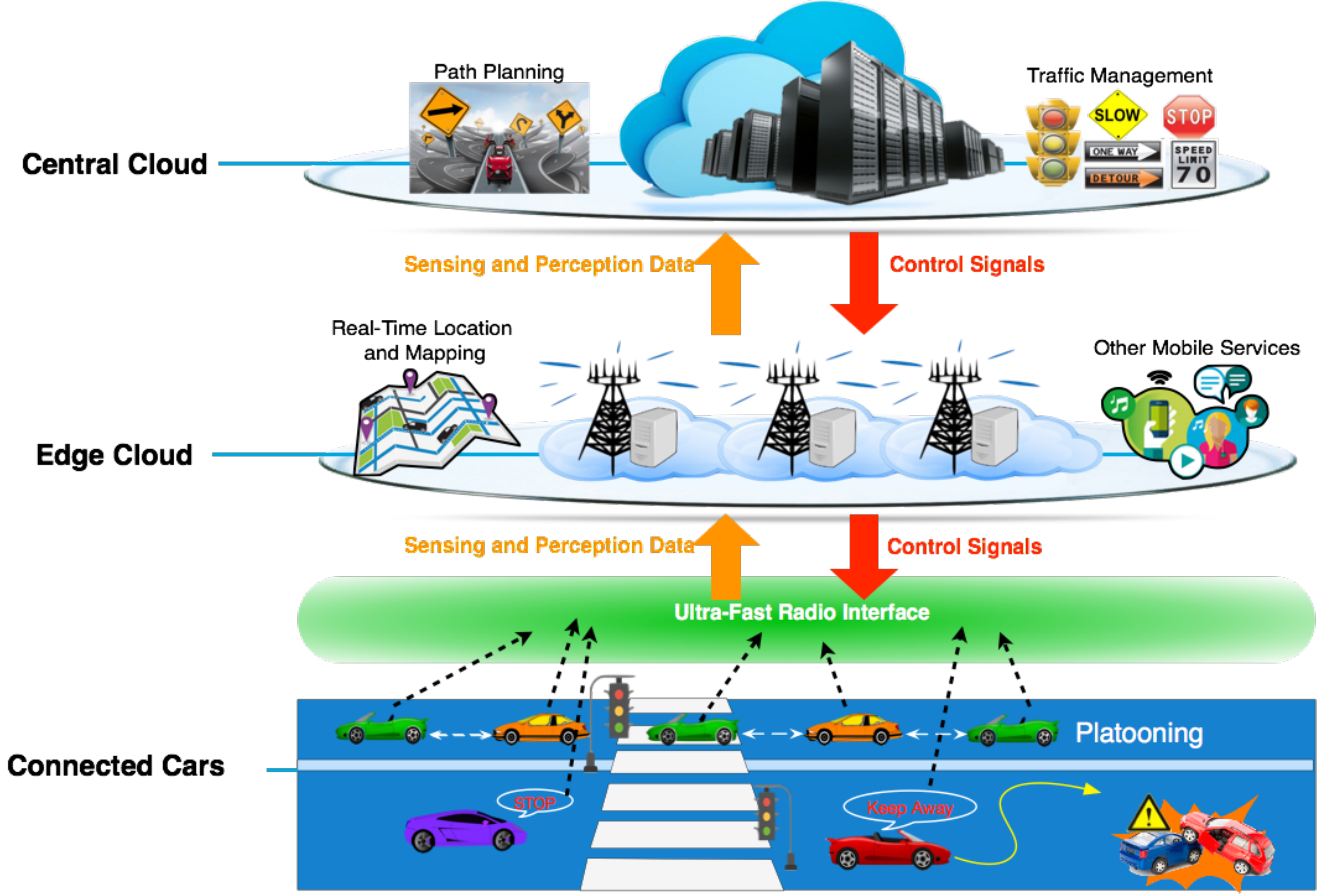}
\end{center}
\caption{\color{black}{MEC for connected vehicles.}}
\label{UseScnCV}
\end{figure*}

The connected vehicle technology can enhance safety, reduce traffic congestion, sense vehicles' behaviors, as well as provide opportunities for numerous value-added services such as the car finder and parking location \cite{Papadimitratos0911,NLu1404,Uhlemann1501}. However, the maturity of such technology is yet to come as the latency requirement cannot be met with the existing connected car clouds, which contributes to an end-to-end latency between 100ms to 1s. MEC is a key enabling technique for connected vehicles by adding computation and geo-distributed services to roadside BSs. By receiving and analyzing the messages from proximate vehicles and roadside sensors, the connected vehicle cloudlets are able to propagate the hazard warnings and latency-sensitive messages within a 20ms end-to-end delay, allowing the drivers to react immediately (as shown in Fig. \ref{UseScnCV}) and make it possible for autonomous driving. The connected vehicle technology has already attracted extensive attention from the automobile manufacturers (e.g., Volvo, Peugeot), automotive supplier (e.g., BOSCH), telecom operators (e.g., Orange, Vodafone, NTT docomo), telecom vendors (e.g., QualComm, Nokia, Huawei), as well as many research institutes. In November 9 2015, Nokia\footnote{\url{https://networks.nokia.com/solutions/mobile-edge-computing}} presented two use cases for connected vehicles on an automotive driving testbed, including the emergency brake light and cooperative passing assistance.

{\color{black}
In addition to connected vehicle systems with automobiles, MEC will also be applicable for enabling connected \emph{unmanned aerial vehicles} (UAVs), which play an increasingly important role in various scenarios such as photography, disaster response, inspection and monitoring, precision agriculture, etc. In 2016, Nokia proposed the \emph{UAV traffic management} (UTM) based MEC architecture for connected UAVs in \cite{UTMinfrastSlides16}, where the UTM unit provides functions of fleet management, automated UAV missions, 3D navigation, and collision avoidance. However, as existing mobile networks are mainly designed for users on the ground, UAVs will have very limited connectivity and bandwidth. Therefore, reconfiguring the mobile networks to guarantee the connectivity and low latency between the UAVs and the infrastructure becomes a critical task for designing MEC systems for connected UAVs.
}

Due to limited space, we omit the description of some other interesting application scenarios, such as active device tracking, RAN-aware content optimization, distributed content and \emph{Domain Name System} (DNS) caching, enterprise networks, as well as safe-and-smart cities. Interested readers may refer to the white papers on MEC \cite{ETSI14,Saguna_intel,Juniper} for details.

{\color{black}\subsection{MEC in 5G Standardizations}

The 5G standard is currently under development, which is to enable the connectivity of a broad range of applications with new functionality, characteristics, and requirements \cite{ericssonWhitepaper}. To achieve these visions, the network features and functionality in 5G networks are foreseen to be migrated from hardware to software, thanks to the recent development of SDN and NFV techniques. Since 2015, MEC (together with SDN and VFN) is recognized by the European \emph{5G infrastructure Public Private Partnership} (5GPPP) research body as one of the key emerging technologies for 5G networks as it is a natural development in the evolution of mobile BSs and the convergence of IT and telecommunication networking \cite{hu2015mobile}. In April 2017, 3GPP has included \emph{supporting edge computing} as one of the high level features in 5G systems in the technical specification document \cite{3GPPTS23501}, which will be introduced in this subsection. We have also identified some innovative features of 5G systems, which would pave the way for the realization, standardization and commercialization of MEC.

\subsubsection{Functionality Supports Offered by 5G Networks} From the 5G network operators' point of view, reducing the end-to-end latency and load on the transport networks are two dominant design targets, which could possibly be achieved with MEC as operators and third part applications could be hosted close to the \emph{user equipment's} (UE's) associated wireless AP. To integrate MEC in 5G systems, the recent 5G technical specifications have explicitly pointed out  necessary functionality supports that should be offered by 5G networks for edge computing, as listed below:
\begin{itemize}
\item The 5G core network should select the traffic to be routed to the applications in the local data networks.

\item {\color{black}{The 5G core network selects a \emph{user plane function} (UPF)
in proximity to the UE to route and execute the traffic steering from the local data networks via the interface, which should be based on the UE's subscription data, UE location, and the data from the \emph{application function} (AF)}}.

\item The 5G network should guarantee the session and service continuity to enable UE and application mobility.

\item The 5G core network and AF should provide information to each other via the \emph{network exposure function} (NEF)\footnote{\color{black}The NEF supports external exposure of capabilities of network functions, which can be categorized into monitoring capability, provisioning capability, and policy/charging capability \cite{3GPPTS23501}.}.

\item The \emph{policy control function} (PCF)\footnote{\color{black}The PCF was defined as a stand-alone functional part of the 5G core network that allows to shape the network behaviour based on the operator policies \cite{PCF5G}.} provides rules for QoS control and charging for the traffic routed to the local data network.
\end{itemize}

\subsubsection{Innovative Features in 5G to Facilitate MEC}

Compared to previous generations of wireless networks, 5G networks possess various innovative features that are beneficial to the realization, standardization, and commercialization of MEC.  Three of them will be detailed in this subsection, including the \emph{support service requirement}, \emph{mobility management strategy}, and \emph{capability of network slicing}.

\begin{itemize}
\item {\textbf{Support Service Requirement:}} In 5G systems, the QoS characteristics (in terms of resource type, priority level, packet delay budget, and packet error rate), which describe the packet forwarding treatment that a QoS flow receives edge-to-edge between the UE and the UPF, are associated with the \emph{5G QoS Indicator} (5QI). In \cite{3GPPTS23501}, a standardized 5QI to QoS mapping table is provided, showing a broad range of services that can be supported in 5G systems. In particular, 5G systems are able to cater the requirements of latency-sensitive applications (e.g., real-time gaming and \emph{vehicular-to-everything} (V2X) messages, which have a stringent packet budget delay requirement, i.e., $<$50ms, and a relatively small packet error rate $<10^{-3}$), and mission-critical services (e.g., push-to-talk signaling that has both low delay ($<$60ms) and small packet error rate ($<10^{-6}$) requirements). These applications coincide with typical MEC applications as mentioned in Section \ref{UseScenarios}, i.e., 5G network is a viable choice for wireless communications in MEC systems.
\item{\textbf{Advanced Mobility Management Strategy:}} {\color{black}{The concept of \emph{mobility pattern} was introduced for designing mobility management strategy for 5G systems. Such strategies may be used by the 5G core network to characterize and optimize  UE mobility}}. Specifically, the mobility pattern could be determined, monitored, and updated by the 5G core network based on the subscription of the UE, statistics of UE mobility, network local policy, and UE assisted information \cite{3GPPTS23501}. The mobility pattern not only plays a central role on designing advanced transmission schemes in wireless communication systems, but also becomes a non-negligible design consideration for many MEC applications discussed in Section \ref{UseScenarios}, e.g., the AR services and connected vehicular applications. Thus, integration of advanced mobility management strategies that make full use of the mobility pattern in 5G network can help to develop an efficient wireless interface for MEC systems. Besides, the mobility pattern obtained from the 5G core network can be further leveraged to design joint radio-and-computational resource management strategies for MEC systems.
\item{\textbf{Capability of Network Slicing:}} Network slicing is a form of agile and virtual network architecture that allows multiple network instances to be created on top of a common shared physical infrastructure\footnote{\color{black}\url{https://5g.co.uk/guides/what-is-network-slicing/}}. Each of the network instances is optimized for a specific service, enabling resource isolation and customized network operations \cite{NGMNslicing}. Due to the heterogeneous types of services that 5G systems need to support (different requirements in terms of functionality and performance), network slicing is regarded as an indispensable feature in 5G systems to support different services running across a single radio access network. Existing studies found that network slicing is of supreme need for three use scenarios, including \emph{ultra-reliable and low latency communication} (URLLC), massive machine type communication (mMTC), and enhanced mobile broadband (eMBB) \cite{NetworkSlicingSlides}. With the capability of network slicing in 5G systems, MEC applications could be provisioned with optimized and dedicated network resources, which could help to reduce the latency incurred by the access networks substantially and support intense access of MEC service subscribers.
\end{itemize}
}

\section{Conclusion}\label{Section:Conclusion}

MEC is an innovative network paradigm to cater for the unprecedented growth of computation demands and the ever-increasing computation quality of user experience requirements. It aims at enabling Cloud Computing capabilities and IT services in close proximity to end users, by pushing abundant computational and storage resources towards the network edges. The direct interaction between mobile devices and edge servers through wireless communications brings the possibility of supporting applications with ultra-low latency requirement, prolonging device battery lives and facilitating highly-efficient network operations. However, they come along with various new design considerations and unique challenges due to reasons such as the complex wireless environments and the inherent limited computation capacities of MEC servers.

In this survey, we presented a comprehensive overview and research outlook of MEC from the communication perspective. To this end, we first summarized the modeling methodologies on key components of MEC systems such as the computation tasks, communications, as well as mobile devices and MEC servers computation. This help characterize the latency and energy performance of MEC systems. Based upon the system modeling, we conducted a comprehensive literature review on recent research efforts on resource management for MEC under various system architectures, which exploit the concepts of computation offloading, joint radio-and-computational resource allocation, MEC server scheduling, as well as multi-server selection and cooperation. A number of potential research directions were then identified, including MEC deployment issues, cache-enabled MEC, mobility management for MEC, green MEC, as well as security-and-privacy issues in MEC. Key research problems and preliminary solutions for each of these directions were elaborated. Finally, we introduced the recent standardization efforts from industry, along with several typical use scenarios. The comprehensive overview and research outlook on MEC provided in this survey hopefully can serve as useful references and valuable guidelines for further in-depth investigations of MEC.


\end{document}